%% file: main.tex
\documentclass[acmsmall, screen, nonacm]{acmart}

\usepackage{booktabs}
\usepackage{multirow}
\usepackage{subcaption}
\usepackage{algorithm}
\usepackage{algpseudocode}

\setcopyright{cc}
\setcctype{by}

\begin{document}

\title{Simulate Before You Scale: High-Fidelity LLM Inference Planning Without GPUs}
\thanks{This is the preprint version of our paper accepted at ACM SIGMETRICS 2027.}

\author{Dipanwita Guhathakurta}
\authornote{These authors contributed equally to this work.}
\orcid{0000-0001-7850-0345}
\affiliation{%
  \institution{IBM Research}
  \city{Yorktown Heights}
  \state{NY}
  \country{USA}                   
}
\email{dipanwita.guhathakurta3@ibm.com}

\author{Mert Toslali}
\authornotemark[1]
\authornote{Corresponding author.}
\orcid{0009-0007-6257-8288}
\affiliation{%
  \institution{IBM Research}
  \city{Cambridge}
  \state{MA}
  \country{USA}                   
}
\email{toslali@ibm.com}

\author{Srinivasan Parthasarathy}
\authornotemark[1]
\orcid{0009-0005-8203-1242}
\affiliation{%
  \institution{IBM Research}
  \city{Yorktown Heights}
  \state{NY}
  \country{USA}                   
}
\email{spartha@us.ibm.com}

\author{Jing Chen}
\orcid{0009-0004-7023-4010}
\affiliation{%
  \institution{IBM Research}
  \city{Yorktown Heights}
  \state{NY}
  \country{USA}                   
}
\email{Jing.Chen2@ibm.com}

\author{Michael Kalantar}
\orcid{0009-0004-1006-5008}
\affiliation{%
  \institution{IBM Research}
  \city{Yorktown Heights}
  \state{NY}
  \country{USA}                   
}
\email{kalantar@us.ibm.com}

\author{Nicholas A. Masluk}
\orcid{0000-0002-7069-4744}
\affiliation{%
  \institution{IBM Research}
  \city{Yorktown Heights}
  \state{NY}
  \country{USA}                   
}
\email{namasluk@us.ibm.com}

\author{Vishakha Ramani}
\orcid{0009-0008-1629-1734}
\affiliation{%
  \institution{IBM Research}
  \city{Yorktown Heights}
  \state{NY}
  \country{USA}                   
}
\email{vishakharamani@ibm.com}

\author{Asser Tantawi}
\orcid{0000-0001-6598-8863}
\affiliation{%
  \institution{IBM Research}
  \city{Yorktown Heights}
  \state{NY}
  \country{USA}                   
}
\email{tantawi@us.ibm.com}

\author{Fabio Oliveira}
\orcid{0009-0001-4918-5208}
\affiliation{%
  \institution{IBM Research}
  \city{Yorktown Heights}
  \state{NY}
  \country{USA}                   
}
\email{fabolive@us.ibm.com}

\renewcommand{\shortauthors}{Guhathakurta et al.}

\begin{abstract}
Deploying large language model (LLM) inference at scale requires jointly selecting hardware layout, parallelism strategy, scheduling policy, admission control, and request routing to meet latency and throughput targets.
These choices interact: a routing decision that improves cache locality can concentrate load, and an admission threshold that triggers only at saturation allows queues to grow deep enough to degrade critical-tier traffic.
Evaluating candidates on GPU clusters is too expensive for iterative exploration, yet existing predictive tools either ignore batching and queueing dynamics or model only fragments of the pipeline.
In this paper, we present BLIS, a deterministic, CPU-only discrete-event simulator that models the full inference serving pipeline (admission, routing, scheduling, prefill/decode execution, KV-cache management, and preemption).
BLIS pairs analytical basis functions with learned scaling coefficients and overhead terms, fit once globally, to predict step latency without per-model or per-GPU profiling.
Across 36 experiments spanning six models and three GPU types, BLIS achieves 6.7\% median end-to-end error while running around 200$\times$ faster than real execution.
In a configuration search over 420 serving configurations for Llama-2-70B, BLIS identifies a 4-GPU deployment that meets a 300\,ms mean TTFT service-level objective (SLO), whereas competing tools either require double the GPUs or recommend configurations that fail under load.
Two policies evolved entirely in simulation were validated on real hardware and merged into llm-d, a widely adopted open-source platform for distributed LLM serving: an admission policy that reduces critical-tier TTFT P90 by up to 82\%, and a flow-control policy that reduces E2E P50 latency by 32--48\% on overloaded workloads.
\end{abstract}

\ccsdesc[500]{Computing methodologies~Discrete-event simulation}
\ccsdesc[500]{General and reference~Performance}
\ccsdesc[500]{Computing methodologies~Modeling and simulation}

\keywords{LLM inference serving; latency modeling; performance prediction; configuration search; admission control; tensor parallelism; KV-cache management}

\maketitle

\section{Introduction}
\label{sec:intro}

Deploying a large language model (LLM) inference platform requires choosing routing policies, admission behavior, batching settings, and KV-cache reuse strategies under concrete time-to-first-token (TTFT), inter-token latency (ITL), throughput, and cost constraints~\cite{kwon2023vllm,zhong2024distserve}.
These choices are coupled.
A routing policy that improves prefix-cache hit rates can concentrate load on a subset of instances; a chunk size that helps one workload can hurt another; an admission policy that sheds load too late lets queues grow deep enough to degrade the traffic it protects.
Teams must predict these coupled effects before committing GPU capacity or promoting a change to production.
Confirming these tradeoffs reliably means measuring them on a GPU cluster: one workload evaluation (e.g., 10{,}000 requests at 10 requests per second (rps) served by Llama-2-70B on NVIDIA H100~\cite{nvidia2024h100datasheet} nodes) takes roughly 18 minutes of sustained GPU time, plus provisioning, download, and warm-up. A meaningful exploration repeats this across hundreds of configurations, policies, and workloads, so the aggregate cost, not any single run, makes broad exploration impractical.

\begin{figure}[tbp]
  \centering
  \includegraphics[width=0.7\columnwidth]{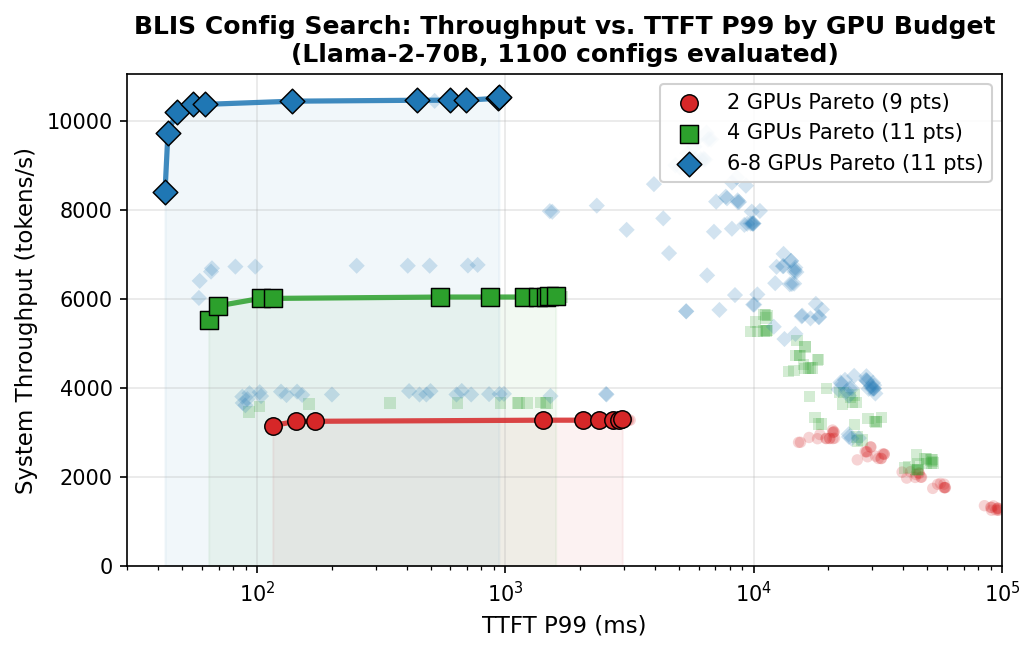}
  \caption{BLIS-only multi-objective configuration search over 1,100 serving configurations for Llama-2-70B on H100 GPUs, evaluated in minutes on a CPU. Each point is one configuration; bold markers trace the Pareto frontier per GPU budget tier (2, 4, and 6--8 GPUs). The search jointly optimizes throughput and TTFT P99 across arrival-rate sweeps. The four-tool comparison in \S\ref{sec:config} uses a 420-configuration subset of this space with mean TTFT as the shared metric.}
  \label{fig:pareto}
\end{figure}

A simulator that faithfully reproduces these coupled dynamics without requiring GPUs is useful in two scenarios.
The first is \emph{configuration search}: sweeping hardware budgets and serving-stack parameters to identify configurations worth deploying on real hardware.
Figure~\ref{fig:pareto} illustrates: BLIS drives a multi-objective search over 1,100 candidate configurations for Llama-2-70B on H100 GPUs, jointly optimizing throughput and TTFT P99 across three GPU budget tiers (2, 4, and 6--8 GPUs) with arrival-rate sweeps. The four-tool comparison in \S\ref{sec:config} uses a 420-configuration subset of this space with mean TTFT as the shared metric, since not all tools predict tail percentiles.
By simulating per-request queueing and scheduling, BLIS predicts tail-latency percentiles (P90, P99) that closed-form estimators cannot (validated against ground truth in \S\ref{subsec:tail}).
The search reveals that throughput plateaus well before tail latency degrades, and that the optimal operating point differs by tier.

The second is \emph{policy evolution}: exploring routing, admission, and batching alternatives before investing the engineering effort to implement and validate candidates on a GPU cluster.
Evaluating policy variants (an admission curve that sheds load under saturation, a routing weight that rewards prefix-cache reuse, a chunk size that governs prefill/decode interleaving) across workload suites is routine in serving-system development, and each evaluation on a GPU cluster takes hours.
With a simulator, broad exploration happens locally on a CPU, and cluster validation is reserved for the small set of policies or configurations that simulation identifies as worth the cluster time.

Existing tools occupy different points in an accuracy, cost, and coverage tradeoff, and each addresses a subset of this problem (\S\ref{sec:background}).
Benchmarks such as inference-perf~\cite{inferenceperf2024} and GuideLLM~\cite{guidellm2024} run the real system and define ground truth, but evaluate one configuration at a time, so exploration cost scales linearly with the number of candidates.
Closed-form estimators such as llm-optimizer~\cite{bentoml2024llmoptimizer} predict without profiling but ignore continuous batching, KV-cache pressure, and queueing, mispredicting by 20--90\% under load; execution-modeling tools such as Vidur~\cite{agrawal2024vidur}, AIConfigurator~\cite{xu2026aiconfigurator}, and LLMServingSim~\cite{cho2026llmservingsim2} model execution more faithfully (median error 17--619\%, highest outside the regime each was calibrated for) but depend on profiling and cannot cover hardware absent from their profiled set without a fresh calibration run.
These tools leave two gaps, which motivate our two technical contributions.
First, accuracy and generalization are in tension: predicting a new model or GPU without re-profiling requires a latency model derived from architectural and hardware specifications, with coefficients fit once, not re-fit per configuration.
Second, the stages of the serving pipeline (admission, batching, queueing) are coupled: each shapes the load the others see. A tool that models them in isolation, or reasons only about steady-state averages, cannot reproduce the load-dependent dynamics (\S\ref{sec:coupling}); capturing them at speed without GPUs requires simulating the pipeline end to end.

We present BLIS, a deterministic, CPU-only discrete-event simulator~\cite{banks2010des} that addresses this problem through three contributions:
\begin{enumerate}
  \item \textbf{Concept.} BLIS introduces a latency model combining roofline-inspired~\cite{williams2009roofline} basis functions with learned, architecture-aware parameters. The model is fit once across a set of models, GPUs, and tensor-parallelism (TP) configurations; new models (dense or mixture-of-experts), new GPU types and new TP configurations are then predicted without re-profiling. (\S\ref{sec:design})
  \item \textbf{System.} BLIS realizes a full-pipeline, coupled discrete-event simulation spanning admission, routing, scheduling, prefill/decode execution, KV-cache management, and preemption, with a validation methodology for systematic comparison against real-cluster measurements. (\S\ref{sec:eval})
  \item \textbf{Evidence.} We show that BLIS is accurate and fast:
  \begin{itemize}
    \item \emph{Fidelity:} BLIS reproduces measured latency within 6.7\% median error across 36 experiments (six models, three GPU types, multiple configurations), nearly 200$\times$ faster than real execution. (\S\ref{sec:eval})
    \item \emph{Configuration search:} BLIS explores hundreds of configurations in minutes to identify candidate Pareto-optimal deployments per GPU budget tier. We validate top recommendations from all four tools (BLIS and three baselines: Vidur, AIConfigurator, and LLM-Optimizer) on an 8$\times$H100 cluster; BLIS's recommendations meet the latency SLO while a closed-form estimator's configuration fails catastrophically under load. (\S\ref{sec:config})
    \item \emph{Real-world impact:} Two policies evolved entirely in simulation were validated on real hardware and merged upstream into llm-d\footnote{\url{https://llm-d.ai}}: an admission policy that reduces critical-tier TTFT P90 by up to 82\% under overload, and a flow-control policy that reduces E2E P50 by 32--48\% on overloaded workloads by proportionally throttling lower-priority traffic. (\S\ref{sec:evolution})
  \end{itemize}
\end{enumerate}

\noindent BLIS's accuracy comes from pairing these first two contributions. An architecture- and hardware-aware latency model, whose coefficients are shared and fit once across 15 real-system experiments, drives a full-pipeline discrete-event simulation of the coupled queueing and batching dynamics. Together they capture both hardware-aware step latency and load-dependent serving behavior, whereas closed-form estimators predict latency without simulating how batch composition and queue depth evolve under load, and profiled-operator simulators reproduce those dynamics but require a calibration run per new model, GPU, or configuration (\S\ref{subsec:fitting}).

\noindent BLIS is open-source, with documentation at \url{https://inference-sim.github.io/inference-sim/latest/}; the code is available at \url{https://github.com/inference-sim/inference-sim/releases/tag/v0.9.2}.

\input{background}

\input{design}

\input{experiments_accuracy}

\input{config_search}

\section{Policy Evolution}
\label{sec:evolution}

Beyond predicting a deployment's latency, BLIS enables \emph{policy evolution}: iterating toward better serving policies by evaluating candidates in simulation before committing GPU time.
We demonstrate this with two case studies, an admission-control and a flow-control policy, each evolved and evaluated entirely in simulation, validated on a real cluster, and merged upstream into the open-source serving platform~\cite{llmd2024}.

\input{case_study}

\input{limitations}

\section{Related Work}
\label{sec:related}

\smallskip\noindent\textbf{Architectural and hardware simulators.}
Cycle-level simulators such as LPGSim~\cite{nam2026lpgsim} and ACALSim~\cite{lin2026acalsim} model GPU microarchitecture, and LLMCompass~\cite{zhang2024llmcompass} extends this to accelerator \emph{design}; BLIS instead abstracts per-GPU execution with a roofline model~\cite{williams2009roofline} and advances through discrete events~\cite{banks2010des} where the coupled dynamics of \S\ref{sec:coupling} arise.
Learned predictors such as NeuSight~\cite{lee2025neusight} likewise generalize to unseen GPUs but forecast operator latency in isolation, whereas BLIS embeds its estimate in a coupled simulation of end-to-end behavior under contention.

\smallskip\noindent\textbf{LLM-serving simulators and analytical predictors.}
The tools closest to BLIS predict per-deployment latency: discrete-event simulators (Vidur~\cite{agrawal2024vidur}, LLMServingSim~\cite{cho2024llmservingsim,cho2026llmservingsim2}), the steady-state estimator AIConfigurator~\cite{xu2026aiconfigurator}, and closed-form estimators (llm-optimizer~\cite{bentoml2024llmoptimizer}, GenZ~\cite{bambhaniya2024genz}); \S\ref{sec:existing} contrasts their scope with BLIS and \S\ref{sec:eval} their accuracy.
Others plan configuration by searching parallelism~\cite{lin2024apex}, composing per-kernel latencies~\cite{infersim2025}, or measuring candidates~\cite{lazuka2024llmpilot}; a trace-driven simulator characterizes prefix-cache reuse~\cite{2026unicache}.

\smallskip\noindent\textbf{Serving and scheduling policies.}
Recent systems refine individual control axes: preemption granularity, phase-aware scheduling, and learned routing~\cite{hsieh2026flowprefill,cho2026pascal,lim2026lodestar}; energy-aware frequency control, online/offline co-serving, and network-layer flow scheduling~\cite{yu2025voltana,qiao2024conserve,sun2026mfs}; and cluster placement across heterogeneous GPUs cast as max-flow~\cite{mei2025helix} or mixed-integer planning~\cite{cheng2026milp}.
BLIS forms a substrate for evaluating policies through pluggable interfaces (\S\ref{subsec:policy}) without kernel changes.

\section{Conclusion}
\label{sec:conclusion}

We presented BLIS, a deterministic, CPU-only discrete-event simulator that predicts the latency and throughput of an LLM serving deployment without per-configuration profiling.
BLIS pairs a physics-informed latency model, fit once, with a coupled simulation of the full serving pipeline.

Across 36 experiments spanning six models, three GPU types, and four workloads, BLIS predicts end-to-end latency to a median 6.7\% error and runs roughly 200$\times$ faster than real execution, generalizing across model families and GPUs without re-profiling.
Driving a configuration search for Llama-2-70B on H100, BLIS identifies a 4-GPU deployment that holds a 300\,ms mean TTFT SLO under sustained load, and real-cluster runs confirm the validity of this configuration.
Beyond benchmarking, an admission policy evolved in simulation was merged into the open-source llm-d serving stack, where it reduced critical-tier TTFT P90 by up to 82\% under overload.
A complementary flow-control policy, evolved the same way and likewise merged into llm-d, throttles rather than sheds lower-priority traffic, reducing end-to-end P50 latency by 32--48\% on overloaded workloads.

More broadly, by predicting latency on a CPU in seconds and exposing admission, routing, and scheduling as replaceable interfaces under a deterministic fitness function, BLIS enables broad configuration search and policy evolution, reducing reliance on cluster-hours benchmarking and pointing toward serving policies designed by exploration rather than by hand.

\bibliographystyle{ACM-Reference-Format}
\bibliography{references}

\appendix

\section{Accuracy Across Config Sweep}
\label{app:config-sweep}

\begin{figure}[tbp]
  \centering
  \includegraphics[width=\columnwidth]{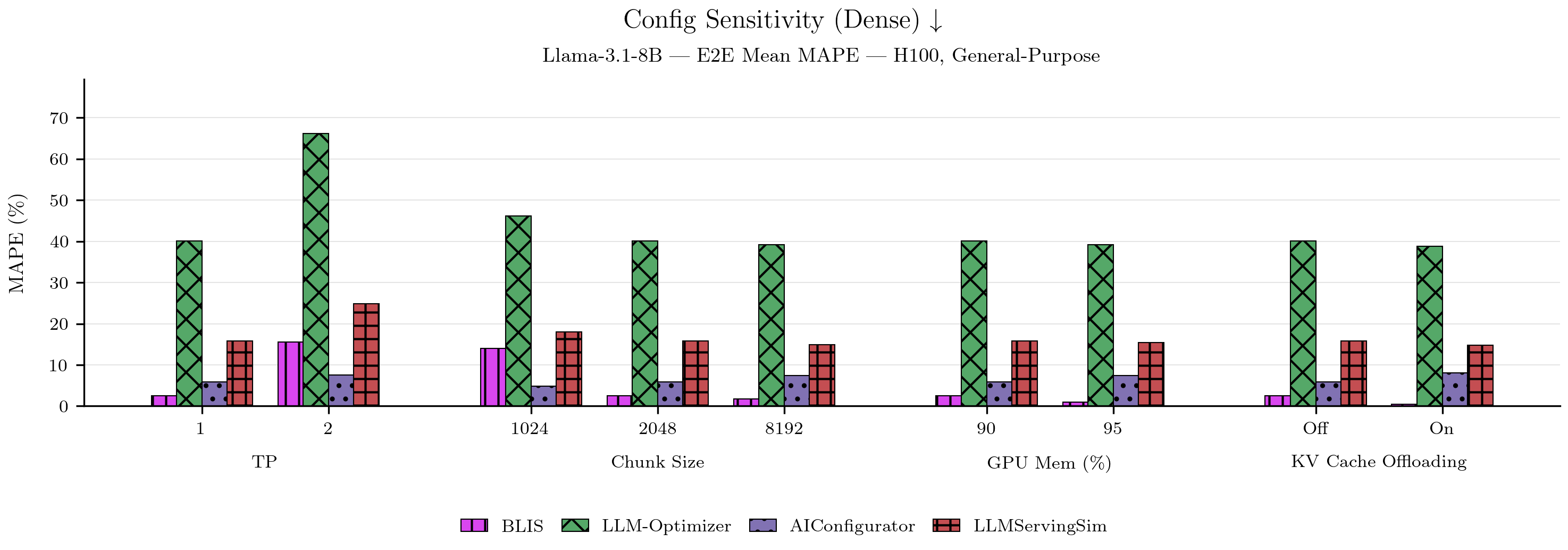}\\
  \includegraphics[width=\columnwidth]{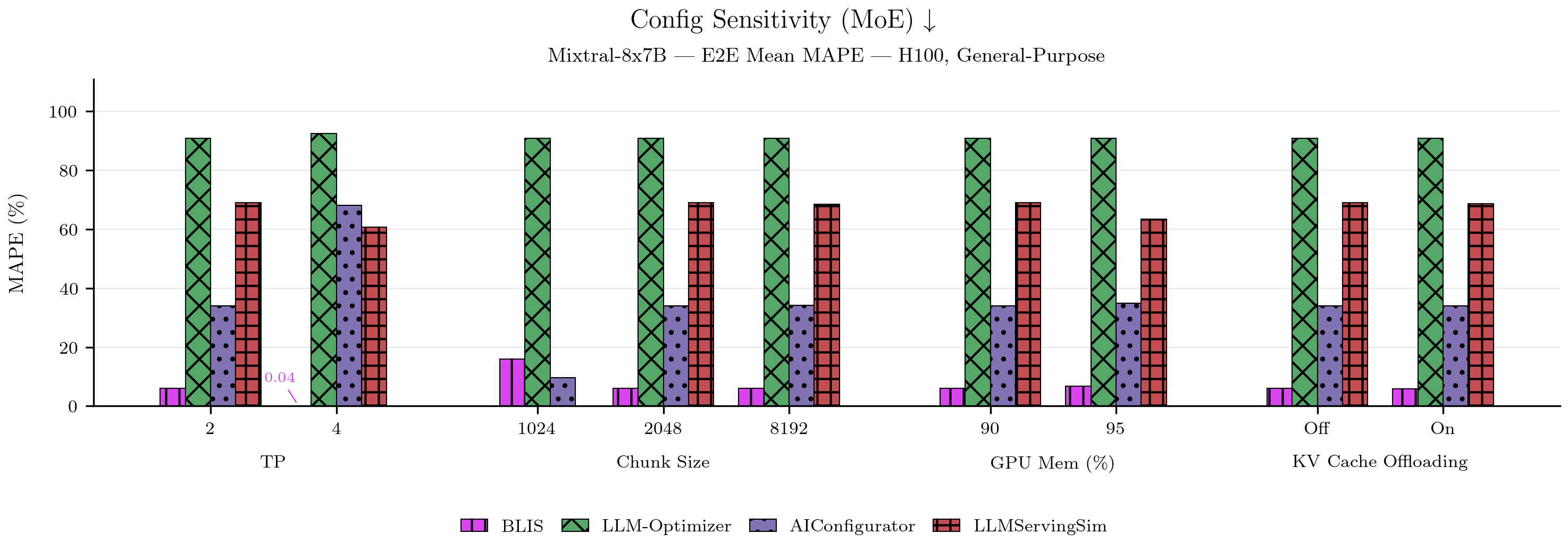}
  \caption{Per-tool E2E MAPE under single-parameter configuration sweeps on H100 (\texttt{mbt}=2048, \texttt{gpu\_mem}=0.9, offloading off). Top: Llama-3.1-8B (dense). Bottom: Mixtral-8x7B (MoE).}
  \label{fig:config-sensitivity}
\end{figure}

This appendix expands the configuration-sweep summary of \S\ref{subsec:headtohead}.
Figure~\ref{fig:config-sensitivity} reports E2E MAPE under controlled single-parameter sweeps on H100 with a general-purpose workload.
The top panel uses Llama-3.1-8B (dense) and the bottom panel uses Mixtral-8x7B (MoE).
Each group varies one serving parameter (TP degree, chunk size, GPU memory utilization, or KV-cache offloading) while holding all others at baseline.

BLIS tracks the ground truth closely at baseline configurations; on the dense model, error increases with tensor-parallel degree (2.5\% at TP=1 to 15.5\% at TP=2) because BLIS estimates all-reduce latency from NVLink bandwidth and message size rather than measured communication traces, and this estimate under-accounts for contention when concurrent batches compete for interconnect bandwidth.
At smaller chunk sizes, more frequent prefill/decode interleaving amplifies scheduling-order sensitivity.
LLM-Optimizer and AIConfigurator do not accept chunk size, offloading, or memory utilization as inputs, so their predictions remain constant across those sweeps: they predict the same latency regardless of how the server is configured, which is correct only at the default setting.
For the MoE panel, LLM-Optimizer treats expert routing as dense, overestimating compute and yielding high MAPE.
BLIS and AIConfigurator model expert routing explicitly and maintain lower error.

\section{Simulation Runtime}
\label{app:runtime}

Table~\ref{tab:accuracy-runtime} reports per-tool median simulation runtime and speedup over real-hardware execution, supplementing the accuracy-speed tradeoff of \S\ref{subsec:costaccuracy}.

\begin{table}[htbp]
  \centering
  \caption{Median simulation runtime and speedup over real-hardware execution across all experiments. Speedup is the median of per-experiment ratios to real execution; LLMServingSim is evaluated on a 1{,}000-request subset (\S\ref{subsec:setup}), so its runtime and speedup are not directly comparable to the full-trace tools.}
  \label{tab:accuracy-runtime}
  \input{accuracy_runtime.tex}
\end{table}

\section{Concurrency Derivation for LLM-Optimizer and AIConfigurator}
\label{app:concurrency}

LLM-Optimizer and AIConfigurator predict per-request latency as a function of \emph{concurrency} (number of in-flight requests) rather than \emph{arrival rate}.
Production workloads are naturally specified in terms of user-facing quantities like request rate, not system internals like concurrency; operators set SLOs on throughput and latency at a given traffic level, not on a fixed number of in-flight requests.
Since our evaluation uses rate-driven workloads to reflect this reality, we derive the appropriate concurrency input for each tool using throughput-matching, a self-consistent procedure that requires no ground-truth latency measurements.

\smallskip\noindent\textbf{Procedure.}
For each experiment, we sweep a geometric series of concurrency levels $c \in \{1,\allowbreak 2,\allowbreak 4,\allowbreak 8,\allowbreak \ldots,\allowbreak 1024\}$ (capped at the vLLM \texttt{max\_num\_seqs} limit).
At each level, the tool returns a predicted throughput $\hat{T}(c)$ in requests per second.
We select the concurrency $c^*$ whose predicted throughput best matches the workload's known arrival rate $\lambda$:
\begin{equation}
  c^* = \arg\min_{c} \left| \frac{\hat{T}(c) - \lambda}{\lambda} \right|
\end{equation}
The latency predictions (TTFT, ITL, E2E) at $c^*$ are then taken as the tool's output for that experiment.

\smallskip\noindent\textbf{AIConfigurator specifics.}
AIConfigurator's SDK returns a Pareto DataFrame spanning multiple concurrency levels in a single call.
We filter to rows matching the experiment's tensor-parallel degree and apply the same throughput-matching procedure on the \texttt{seq/s} column.
E2E latency is derived from the matched row as $\text{E2E} = \text{TTFT} + (\text{time per output token}) \times (N_{\text{out}} - 1)$, where $N_{\text{out}}$ is the output sequence length.

\smallskip\noindent\textbf{Matching fidelity.}
Because both tools produce throughput predictions at discrete concurrency levels, the selected operating point $\hat{T}(c^*)$ does not in general equal $\lambda$ exactly.
We quantify this discretization residual $|\hat{T}(c^*) - \lambda|/\lambda$ at the summary level (duration-weighted average across stages) for all experiments in our evaluation.
LLM-Optimizer's power-of-two concurrency grid yields a mean residual of 8.7\%; AIConfigurator's Pareto grid yields a mean residual of 12.3\%.
In both cases, the matching residual is substantially smaller than the tools' intrinsic prediction error on latency metrics (\S\ref{subsec:headtohead}), confirming that the discretization introduced by throughput-matching is not the dominant source of inaccuracy in these baselines.
The only workload quantity used as input is the arrival rate $\lambda$, which is a \emph{design parameter} of the load profile (set before the run), not a measurement; no ground-truth latency or throughput measurements enter the concurrency derivation.

\section{Why Re-Profiling Vidur Is Insufficient}
\label{app:vidur}

A natural question is whether Vidur's high TTFT error (\S\ref{subsec:headtohead}) could be resolved by collecting fresh profiling data.
Re-profiling addresses the wrong problem: Vidur's simulator models a system whose scheduling semantics, state representation, and execution abstraction are structurally incompatible with modern vLLM.
Fresh timing data fed into the existing simulator still produces predictions for the wrong system.
Table~\ref{tab:vidur-gaps} enumerates the gaps; we highlight the three most consequential below.

\smallskip\noindent\textbf{Monolithic prefill.}
Vidur models prefill as all-or-nothing (entire prompt in one pass) with a binary prefill/decode flag.
vLLM~v1 chunks prefill dynamically via a shared token budget with concurrent decode requests, requiring partial-progress state that Vidur's request representation cannot express.

\smallskip\noindent\textbf{Immutable batches.}
Vidur's batches are fixed once formed; vLLM~v1 rebuilds the batch at every iteration with mid-execution preemption.
The preemption cascades and batch-composition dynamics that determine tail latency under load have no analog in Vidur's event model.

\smallskip\noindent\textbf{Incompatible profiling infrastructure.}
Vidur's profiling imports from \texttt{sarathi} (a pre-chunked-prefill fork) and produces data in a schema (monolithic prefill latency, fixed-batch decode latency) incompatible with the heterogeneous mixed-batch forward passes vLLM~v1 executes.
Replacing the profiling target does not help: the predictor expects per-phase latencies for batch sizes that no longer correspond to how vLLM forms batches.

\medskip\noindent
Resolving these gaps would require rewriting the scheduler event model, request state machine, batch formation logic, and profiling infrastructure, which constitutes building a new simulator rather than re-profiling the existing one.

\begin{table}[htbp]
  \centering
  \small
  \caption{Architectural comparison between Vidur's simulator model and production vLLM~v1. Each row represents a structural incompatibility that cannot be resolved by collecting new profiling data.}
  \label{tab:vidur-gaps}
  \begin{tabular}{@{}lll@{}}
    \toprule
    \textbf{Feature} & \textbf{Vidur} & \textbf{vLLM v1} \\
    \midrule
    Prefill execution & Monolithic (all-or-nothing) & Dynamic chunked (token budget) \\
    Scheduling granularity & Batch-level (immutable) & Iteration-level (rebuilt each step) \\
    Prefix caching & Not modeled & Content-addressed block hashing \\
    Phase model & Binary prefill/decode flag & Unified token-progress counter \\
    Speculative decoding & Not modeled & Integrated with scheduling \\
    Profiling backend & \texttt{sarathi} (diverged fork) & N/A (vLLM-native) \\
    Attention workload & Homogeneous (batch\_size=1) & Heterogeneous mixed batches \\
    \bottomrule
  \end{tabular}
\end{table}

\section{Roofline MFU Factors}
\label{app:mfu}

The roofline configuration used in the ablation study (\S\ref{subsec:ablations}) scales the analytical basis functions by fixed per-phase, per-GPU Model-FLOPs-Utilization (MFU) factors.
Table~\ref{tab:mfu} lists these values; they are derived from published utilization benchmarks~\cite{shah2024flashattention3, korthikanti2023megatron, reddi2020mlperfinference} and set as constants in the hardware configuration.
Prefill factors exceed decode factors on every GPU because small-batch decode runs at an arithmetic intensity well below the roofline ridge.

\begin{table}[htbp]
  \centering
  \small
  \caption{Fixed MFU factors used by the roofline configuration.}
  \label{tab:mfu}
  \begin{tabular}{@{}lcc@{}}
    \toprule
    \textbf{GPU} & \textbf{Prefill} & \textbf{Decode} \\
    \midrule
    H100 & 0.45 & 0.30 \\
    A100-SXM & 0.38 & 0.18 \\
    L40S & 0.32 & 0.08 \\
    \bottomrule
  \end{tabular}
\end{table}

\section{Latency-Model Coefficients}
\label{app:training}

Section~\ref{subsec:fitting} describes the two-loop fitting procedure, the loss (Equation~\ref{eq:training-loss}), the training set, the inner-loop optimizers, and why the fit transfers to unseen configurations.
Table~\ref{tab:fitted-coeffs} complements that description with the role and unit of every fitted coefficient. The four scaling factors ($\beta_{1a}$--$\beta_{2b}$) absorb the gap between peak and observed throughput per phase, and the additive terms ($\beta_5$--$\beta_8$, $\alpha_0$--$\alpha_2$) carry the fixed overheads the roofline omits.

\begin{table}[htbp]
  \centering
  \small
  \caption{Latency-model coefficients. Dimensionless scaling factors relate observed to analytical throughput; overhead terms are in microseconds.}
  \label{tab:fitted-coeffs}
  \begin{tabular}{@{}lll@{}}
    \toprule
    \textbf{Coefficient} & \textbf{Role} & \textbf{Unit} \\
    \midrule
    \multicolumn{3}{@{}l}{\emph{$\alpha$: per-request CPU/framework overheads}} \\
    $\alpha_0$ & Request ingestion overhead & $\mu$s \\
    $\alpha_1$ & Detokenization overhead & $\mu$s \\
    $\alpha_2$ & Token streaming overhead & $\mu$s/tok \\
    \midrule
    \multicolumn{3}{@{}l}{\emph{$\beta$: per-step GPU execution scaling and overheads}} \\
    $\beta_{1a}$ & Prefill compute scaling & dimensionless \\
    $\beta_{1b}$ & Prefill memory scaling & dimensionless \\
    $\beta_{2a}$ & Decode compute scaling & dimensionless \\
    $\beta_{2b}$ & Decode memory scaling & dimensionless \\
    $\beta_3$ & Weight loading scaling & dimensionless \\
    $\beta_4$ & TP communication scaling & dimensionless \\
    $\beta_5$ & Per-layer overhead & $\mu$s/layer \\
    $\beta_6$ & Per-request overhead & $\mu$s/req \\
    $\beta_7$ & Per-step constant & $\mu$s/step \\
    $\beta_8$ & Per-MoE-layer overhead & $\mu$s/layer \\
    \bottomrule
  \end{tabular}
\end{table}

\section{Evaluation Coverage}
\label{app:coverage}

Table~\ref{tab:coverage} enumerates every configuration validated in the accuracy study (\S\ref{subsec:setup}).
The design combines breadth across models, workloads, and hardware with depth along the configuration knobs: for the two representative models used in the knob sweep (Llama-3.1-8B and Mixtral-8x7B on H100), we vary tensor-parallel degree, chunk size, GPU-memory utilization, and CPU offloading one knob at a time from the baseline, while the remaining models and both non-H100 GPUs are validated at the baseline configuration.
All deployments use FP16 precision.

\begin{table}[tp]
  \centering
  \footnotesize
  \setlength{\tabcolsep}{4pt}
  \caption{Configurations validated in the accuracy study (\S\ref{subsec:setup}), grouped by GPU. Chunk is \texttt{max\_num\_batched\_tokens}; Mem.\ is \texttt{gpu\_memory\_utilization}; Offload is KV-cache CPU offloading. All deployments use FP16 precision.}
  \label{tab:coverage}
  \begin{tabular}{@{}llcccc@{}}
    \toprule
    \textbf{Model} & \textbf{Workload} & \textbf{TP} & \textbf{Chunk} & \textbf{Mem.} & \textbf{Offload} \\
    \midrule
    \multicolumn{6}{@{}l}{\emph{H100-80GB SXM}} \\
    Llama-3.1-8B  & General-Purpose & 1 & 1024 & 0.90 & off \\
    Llama-3.1-8B  & General-Purpose & 1 & 2048 & 0.90 & off \\
    Llama-3.1-8B  & General-Purpose & 1 & 2048 & 0.90 & on  \\
    Llama-3.1-8B  & General-Purpose & 1 & 2048 & 0.95 & off \\
    Llama-3.1-8B  & General-Purpose & 1 & 8192 & 0.90 & off \\
    Llama-3.1-8B  & General-Purpose & 2 & 2048 & 0.90 & off \\
    Llama-3.1-8B  & Code Generation & 1 & 2048 & 0.90 & off \\
    Llama-3.1-8B  & Roleplay        & 1 & 2048 & 0.90 & off \\
    Llama-3.1-8B  & Reasoning       & 1 & 2048 & 0.90 & off \\
    Qwen3-14B     & General-Purpose & 1 & 2048 & 0.90 & off \\
    Qwen3-14B     & Code Generation & 1 & 2048 & 0.90 & off \\
    Qwen3-14B     & Roleplay        & 1 & 2048 & 0.90 & off \\
    CodeLlama-34B & General-Purpose & 2 & 2048 & 0.90 & off \\
    CodeLlama-34B & Code Generation & 2 & 2048 & 0.90 & on  \\
    CodeLlama-34B & Roleplay        & 2 & 2048 & 0.90 & on  \\
    Llama-2-70B   & General-Purpose & 4 & 2048 & 0.90 & off \\
    Llama-2-70B   & Code Generation & 4 & 2048 & 0.90 & on  \\
    Llama-2-70B   & Roleplay        & 4 & 2048 & 0.90 & on  \\
    Mixtral-8x7B  & General-Purpose & 2 & 1024 & 0.90 & off \\
    Mixtral-8x7B  & General-Purpose & 2 & 2048 & 0.90 & off \\
    Mixtral-8x7B  & General-Purpose & 2 & 2048 & 0.90 & on  \\
    Mixtral-8x7B  & General-Purpose & 2 & 2048 & 0.95 & off \\
    Mixtral-8x7B  & General-Purpose & 2 & 8192 & 0.90 & off \\
    Mixtral-8x7B  & General-Purpose & 4 & 2048 & 0.90 & off \\
    Mixtral-8x7B  & Code Generation & 2 & 2048 & 0.90 & on  \\
    Mixtral-8x7B  & Roleplay        & 2 & 2048 & 0.90 & on  \\
    Mixtral-8x22B & General-Purpose & 8 & 2048 & 0.90 & off \\
    Mixtral-8x22B & Code Generation & 8 & 2048 & 0.90 & off \\
    Mixtral-8x22B & Roleplay        & 8 & 2048 & 0.90 & off \\
    Mixtral-8x22B & Reasoning       & 8 & 2048 & 0.90 & off \\
    \midrule
    \multicolumn{6}{@{}l}{\emph{A100-80GB SXM}} \\
    Llama-3.1-8B  & General-Purpose & 1 & 2048 & 0.90 & off \\
    Qwen3-14B     & General-Purpose & 1 & 2048 & 0.90 & off \\
    CodeLlama-34B & General-Purpose & 2 & 2048 & 0.90 & off \\
    Llama-2-70B   & General-Purpose & 4 & 2048 & 0.90 & off \\
    \midrule
    \multicolumn{6}{@{}l}{\emph{L40S (48\,GB, PCIe)}} \\
    Llama-3.1-8B  & General-Purpose & 1 & 2048 & 0.90 & off \\
    Qwen3-14B     & General-Purpose & 1 & 2048 & 0.90 & off \\
    \bottomrule
  \end{tabular}
\end{table}

\section{Request Lifecycle}
\label{app:lifecycle}

\begin{figure}[tbp]
  \centering
  \includegraphics[width=\columnwidth]{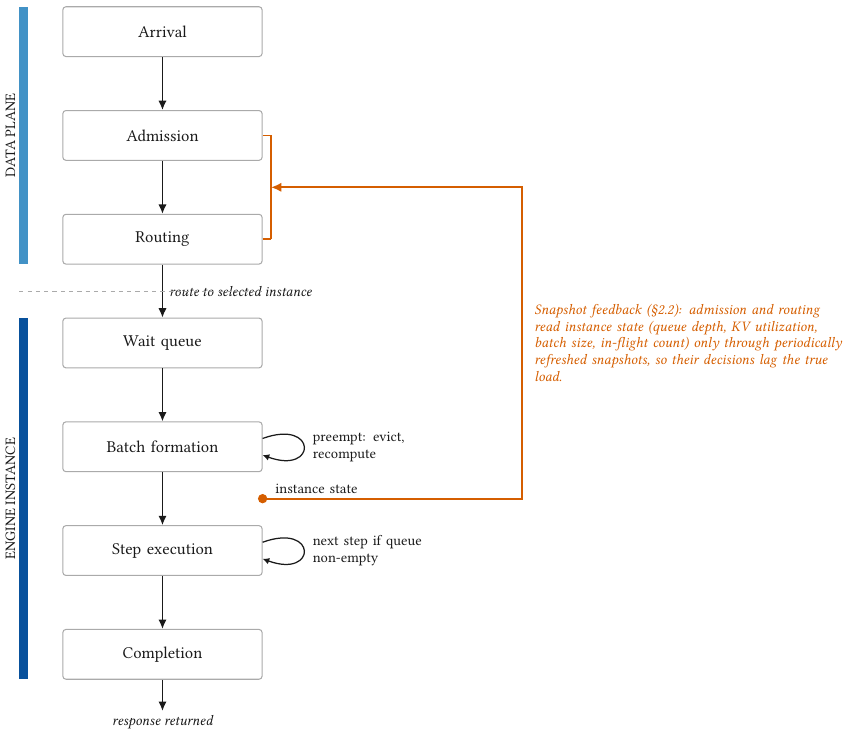}
  \caption{Request lifecycle through the two simulation layers. The back-edge marks the snapshot feedback loop: routing and admission act on periodically refreshed instance state, so their decisions can lag the true load.}
  \label{fig:eventflow}
\end{figure}

Figure~\ref{fig:eventflow} traces a single request through the two simulation layers of \S\ref{subsec:pipeline}, from arrival to completion.
The left bands mark the two scopes: the cluster-level data plane (admission and routing) and the per-instance engine (wait queue, batch formation, step execution).
Two engine-local control loops sit on the spine: batch formation preempts and later recomputes evicted requests under KV-cache pressure, and step execution re-schedules whenever the wait queue is non-empty (the work-conserving invariant).
The vermillion back-edge is the snapshot feedback path of \S\ref{sec:coupling}: admission and routing read instance state only through refreshed snapshots, so their decisions act on stale load.

\section{Configuration Search Details}
\label{app:config-details}

This appendix reports the full search space explored by each tool in \S\ref{sec:config} and the specific configurations deployed for validation on H100 GPUs.

\smallskip\noindent\textbf{Search space.}
Table~\ref{tab:search-space} lists the parameter ranges each tool searched.
BLIS explores the widest space (420 configurations across topology, batching, scheduling, routing, admission, and preemption); the other tools search subsets determined by their modelling scope.

\begin{table}[htbp]
  \centering
  \small
  \caption{Search space per tool. BLIS searches jointly over all listed dimensions via NSGA-II; Vidur sweeps TP and batch size with a fixed vLLM scheduler; AIConfigurator and LLM-Optimizer each produce a fixed grid from their internal logic.}
  \label{tab:search-space}
  \begin{tabular}{@{}llp{5.2cm}r@{}}
    \toprule
    \textbf{Tool} & \textbf{Parameter} & \textbf{Values} & \textbf{Configs} \\
    \midrule
    \multirow{7}{*}{BLIS}
      & TP $\times$ replicas & $\{2,4,8\} \times \{1,2,4\}$ ($\le$8 GPUs) & \multirow{7}{*}{420} \\
      & \texttt{max\_num\_seqs} & \{32, 64, 128, 256, 512\} & \\
      & \texttt{max\_num\_batched\_tokens} & \{2048, 4096, 8192\} & \\
      & Chunked prefill & \{on, off\} & \\
      & Scheduling & \{FCFS, SJF\} & \\
      & Routing & \{round-robin, least-loaded, weighted\} & \\
      & Admission / preemption & \{tier-shed, always-admit\} / \{FCFS, priority\} & \\
    \midrule
    \multirow{3}{*}{Vidur}
      & TP $\times$ replicas & $\{4,8\} \times \{1,2\}$ ($\le$8 GPUs) & \multirow{3}{*}{35} \\
      & \texttt{batch\_size\_cap} & \{32, 64, 128, 256, 512\} & \\
      & Scheduler & \{vllm, sarathi\} (chunk $\in$ \{512, 1024, 2048\}) & \\
    \midrule
    \multirow{2}{*}{AICfg}
      & TP $\times$ replicas & $\{2,4,8\} \times \{1,\ldots,4\}$ ($\le$8 GPUs) & \multirow{2}{*}{13} \\
      & \texttt{max\_num\_seqs} & Tool-determined (20 or 80 or 144) & \\
    \midrule
    \multirow{2}{*}{LLMOpt}
      & TP & $\{2,3,4,5,6,7,8\}$ & \multirow{2}{*}{7} \\
      & \texttt{max\_num\_seqs} & Tool-determined (4, 8, or 16) & \\
    \bottomrule
  \end{tabular}
\end{table}

\smallskip\noindent\textbf{Deployed configurations.}
Table~\ref{tab:deployed-configs} reports the vLLM serving parameters deployed on the H100 cluster for each validated configuration in Table~\ref{tab:validation}.
All deployments share a common baseline: vLLM~v0.22.1, BF16 precision, \texttt{gpu\_memory\_utilization}=0.9, paged KV-cache (block size 16), chunked prefill enabled, and \texttt{max\_model\_len}=4096.
For multi-replica deployments, llm-d routes requests via its Endpoint Picker Plugin (EPP).
Vidur, AIConfigurator, and LLM-Optimizer do not expose routing configuration, so their deployments use the llm-d default EPP profile: weighted scoring over queue-depth (weight~2), KV-cache utilization (weight~2), and prefix-cache hit rate (weight~3).
BLIS searches over routing strategies (\S\ref{sec:config}) but its top-2 Pareto configurations are single-replica, so routing does not apply.
No admission control or flow control was applied (all requests admitted immediately).

\begin{table}[htbp]
  \centering
  \small
  \caption{Deployed vLLM serving parameters for each validated configuration. All use Llama-2-70B on H100 SXM 80\,GB, PP=1, BF16, chunked prefill enabled. ``default'' denotes the vLLM default of 8192 tokens when chunked prefill is on. Single-replica deployments bypass routing.}
  \label{tab:deployed-configs}
  \begin{tabular}{@{}lcccccc@{}}
    \toprule
    \textbf{Tool} & \textbf{TP} & \textbf{Replicas} & \textbf{GPUs} & \texttt{max\_num\_seqs} & \texttt{max\_num\_batched\_tokens} & \textbf{Routing} \\
    \midrule
    BLIS & 4 & 1 & 4 & 64  & 8192 & n/a (1 replica) \\
    BLIS & 4 & 1 & 4 & 256 & 8192 & n/a (1 replica) \\
    \midrule
    Vidur & 4 & 2 & 8 & 64  & default & EPP weighted \\
    Vidur & 4 & 2 & 8 & 256 & default & EPP weighted \\
    \midrule
    AICfg & 2 & 2 & 4 & 20 & 566 & EPP weighted \\
    AICfg & 2 & 3 & 6 & 20 & 566 & EPP weighted \\
    \midrule
    LLMOpt & 2 & 1 & 2 & 4 & default & n/a (1 replica) \\
    LLMOpt & 4 & 1 & 4 & 8 & default & n/a (1 replica) \\
    \bottomrule
  \end{tabular}
\end{table}

\section{Soft-Reflective Ceiling: Per-Workload Results}
\label{app:soft-reflective}

Table~\ref{tab:soft-reflective} reports the full per-workload P50 latency deltas for the soft-reflective ceiling policy (\S\ref{subsec:soft-reflective}).
All measurements are on Qwen3-14B served on 2$\times$H100 GPUs (TP=1), 80\% critical / 20\% sheddable traffic.
Sustainable throughput is estimated at ${\approx}17$ req/s per instance, defined as the arrival rate at which mean E2E latency begins to inflect upward (the knee of the latency-load curve).

\begin{table}[htbp]
  \centering
  \small
  \caption{Soft-reflective ceiling policy: P50 latency change relative to baseline (no policy) by workload profile and arrival rate.}
  \label{tab:soft-reflective}
  \begin{tabular}{@{}lcrrr@{}}
    \toprule
    \textbf{Workload} & \textbf{rps} & \textbf{TTFT $\Delta$} & \textbf{TPOT $\Delta$} & \textbf{E2E $\Delta$} \\
    \midrule
    code\_gen & 4 & ${+}1\%$ & ${-}1\%$ & ${-}1\%$ \\
    code\_gen & 10 & ${+}5\%$ & ${-}1\%$ & ${+}2\%$ \\
    code\_gen & 16 & $\mathbf{{-}35\%}$ & $\mathbf{{-}24\%}$ & $\mathbf{{-}32\%}$ \\
    code\_gen & 24 & ${-}23\%$ & ${-}13\%$ & ${-}22\%$ \\
    \midrule
    chat & 5 & ${+}2\%$ & ${<}1\%$ & ${+}5\%$ \\
    chat & 10 & ${+}3\%$ & ${<}1\%$ & ${+}6\%$ \\
    chat & 25 & ${-}28\%$ & ${-}29\%$ & ${-}24\%$ \\
    chat & 40 & ${-}11\%$ & ${-}19\%$ & ${+}2\%$ \\
    \midrule
    reasoning & 0.2 & ${+}19\%$ & ${+}16\%$ & ${+}31\%$ \\
    reasoning & 0.7 & $\mathbf{{-}85\%}$ & $\mathbf{{-}23\%}$ & $\mathbf{{-}48\%}$ \\
    reasoning & 1.2 & ${+}88\%$ & $\mathbf{{-}32\%}$ & ${+}30\%$ \\
    \bottomrule
  \end{tabular}
\end{table}

\end{document}

%% file: background.tex

\section{Background and Motivation}
\label{sec:background}

Selecting and tuning an LLM serving deployment requires evaluating candidates far faster and more cheaply than real-cluster experiments permit (\S\ref{sec:intro}).
A simulator can occupy this role only if it satisfies three properties at once: \emph{fidelity}, reproducing the latency and throughput operators optimize for; \emph{speed}, evaluating a configuration quickly so it can sit inside a search loop; and \emph{composability}, accepting substitutions of individual policies (admission, routing, scheduling) without internal modification.
Human operators and automated agents rely on the simulator as a deterministic fitness function: identical inputs must yield identical predictions, so an outcome change reflects policy or configuration, not simulation noise (\S\ref{sec:evolution}).
This section establishes the structure of the modeled system (\S\ref{sec:pipeline}), characterizes the coupled dynamics that make its behavior hard to predict (\S\ref{sec:coupling}), reviews gaps in existing tools (\S\ref{sec:existing}), and distills the requirements (\S\ref{sec:requirements}).

\subsection{The LLM Serving Pipeline}
\label{sec:pipeline}

An LLM serving instance generates each response autoregressively in two phases with distinct performance characteristics.
The prefill phase processes the full input prompt in parallel and is compute-bound; the decode phase then produces output tokens one at a time, and because each step reads the entire key-value (KV) cache from GPU memory, it is memory-bound~\cite{kwon2023vllm}.
The KV cache stores the attention state for every token generated so far.
It is allocated in fixed-size blocks and grows with sequence length, so an instance's finite GPU memory bounds the number of requests it can serve concurrently~\cite{kwon2023vllm}.
To raise utilization, modern engines use continuous batching~\cite{yu2022orca}.
A step is one model forward pass over the current batch, producing one token for every decoding request.
All requests in a batch share a single forward pass, so per-step latency is governed by its most expensive operation: a large prefill co-scheduled with many decodes raises the latency of every request in the step.
Chunked prefill, as in Sarathi-Serve~\cite{agrawal2024sarathi}, splits long prefills into smaller units interleaved with decode steps to bound this effect.
When the KV cache saturates, the engine preempts running requests and frees their blocks; a preempted request must later recompute or reload its evicted KV state before it can resume~\cite{kwon2023vllm}.

At cluster scale, a deployment comprises many such instances behind a shared admission and routing layer.
Admission control gates incoming requests, a router selects a target instance for each admitted request (using policies such as load balancing or prefix-aware placement), and per-instance scheduling orders the local wait queue.
Disaggregated deployments further split the prefill and decode phases across separate instance pools so that each can be provisioned and scaled independently~\cite{zhong2024distserve}.
These admission and routing decisions operate on periodic snapshots of instance state (queue depth, KV-cache utilization, in-flight count) that are stale: by the time one is acted upon, the state it describes may already have changed, particularly under shifting load.

\subsection{Why Modeling Is Hard: Coupled Dynamics}
\label{sec:coupling}

This pipeline is hard to predict because its components are coupled: a perturbation in one stage propagates through others via load-dependent feedback.
We group these into three mechanisms.

\smallskip\noindent\textbf{Cross-request coupling.}
Requests in a batch share one forward pass, so a step costs the maximum over its operations, not their average. One long prefill chunk stalls every decode it shares a step with, an effect chunked prefill bounds but does not remove~\cite{agrawal2024sarathi}.
Continuous batching changes this composition every step, so per-request latency depends on the entire arrival pattern, not isolated requests. Steady-state analysis, which assumes a fixed or average batch, misses it.

\smallskip\noindent\textbf{KV-cache cascades.}
Memory pressure couples requests through a feedback loop: when the KV cache fills, the engine preempts running requests~\cite{kwon2023vllm}, and recomputing their evicted state on re-admission adds load that raises pressure further.
One large request entering a nearly full instance can thus cascade into preemptions that degrade many co-located requests; the effect is non-linear in load, absent below the memory limit and growing with the number of evictions beyond it.

\smallskip\noindent\textbf{Stale-snapshot feedback.}
At cluster scale, admission, routing, and flow-control decisions act on the stale snapshots of \S\ref{sec:pipeline}, creating feedback of their own.
When several requests arrive within one snapshot interval, a load-balancing router sends them all to the instance that looks least loaded; the resulting overload raises its latency, yet the stale snapshot keeps attracting traffic until it refreshes, and saturation thresholds lag for the same reason.
Because these loops interact, a deployment's behavior depends on the joint configuration of its policies, not any one alone.

These dynamics determine whether latency targets are met.
The default admission controller in a widely adopted serving platform~\cite{llmd2024} admits all traffic until cluster saturation reaches its limit, then sheds \emph{sheddable}-tier (lower-priority) requests outright; because that threshold fires only once queues are deep and the KV cache nearly full, critical requests admitted just before it suffer elevated TTFT from contention with queued sheddable traffic, a cliff that proactive shedding removes (\S\ref{sec:evolution}).

\subsection{Existing Approaches}
\label{sec:existing}

Existing tools occupy different points in a tradeoff among accuracy, cost, and coverage, and each rests on assumptions that bound which of the dynamics in \S\ref{sec:coupling} it can reproduce.

\smallskip\noindent\textbf{Benchmarks.}
Tools such as inference-perf~\cite{inferenceperf2024} and GuideLLM~\cite{guidellm2024} run the real system and report ground-truth measurements, but they measure rather than predict and exploring an alternative policy requires redeploying the system under test.
They serve as the reference against which predictive tools are validated.

\smallskip\noindent\textbf{Closed-form estimators.}
They predict latency and throughput analytically rather than by simulating execution.
llm-optimizer~\cite{bentoml2024llmoptimizer} includes an estimator that works from hardware specifications, requiring no profiling of the target model or GPU.
Its estimator reasons about isolated requests and static bottlenecks at a fixed, steady-state operating point, and ignores continuous batching, KV-cache pressure, or queueing.

\smallskip\noindent\textbf{Simulators.}
They model the system's execution more directly but differ in both scope and method: \emph{discrete-event} simulators (Vidur, LLMServingSim) advance through individual request-arrival and batch-step events and so reproduce queueing and batch dynamics, whereas \emph{steady-state} estimators (AIConfigurator) compute latency analytically at a fixed operating point without unrolling those events.
Vidur~\cite{agrawal2024vidur} simulates a cluster of replicas behind a global request router and predicts per-operator runtimes with a random-forest model fit per (model, GPU, TP) configuration, so TP communication is absorbed into the profiled operator times rather than simulated explicitly. Its replica scheduler follows Sarathi-Serve~\cite{agrawal2024sarathi}, a vLLM fork whose chunked-prefill scheduling and KV-cache management differ from current production engines; it does not model admission control or prefill/decode disaggregation.
AIConfigurator~\cite{xu2026aiconfigurator} profiles individual operators (GEMM, attention, communication, MoE) per GPU type and precision into a performance database, then composes the interpolated operator times for a given architecture and tensor-parallel degree to estimate end-to-end latency; it explicitly models prefill/decode disaggregation. As a steady-state estimator it computes latency at a fixed concurrency, so it does not capture how batch composition and queue depth evolve under load, and falls back to speed-of-light estimates for hardware absent from the database. InferSim~\cite{infersim2025} takes the same profiled-operator approach and likewise omits cluster-level queueing, so AIConfigurator represents this class in our evaluation.
LLMServingSim~2.0~\cite{cho2026llmservingsim2} extends the original~\cite{cho2024llmservingsim} to heterogeneous and disaggregated infrastructure; we evaluate 2.0 (\S\ref{sec:eval}).
Both model multi-instance serving with profiled kernel timings, request routing, and KV-cache pressure handled by swapping blocks to host memory, and, unlike Vidur, simulate inter-GPU communication and its overlap with compute explicitly rather than folding it into the kernel profiles; neither models admission control or queue-full rejection.
Each tool captures a subset of the pipeline; in our evaluation their median end-to-end error spans 17\% to 619\% (\S\ref{sec:eval}), reaching the upper end outside the regime a tool was calibrated for.
Vidur and LLMServingSim require re-profiling for each new model, GPU, or TP, restricting coverage to already-profiled configurations; AIConfigurator profiles only per GPU type and precision, deriving unseen models and parallelism degrees analytically, but cannot cover a GPU absent from its database.

\begin{table}[tbp]
  \centering
  \caption{Predictive approaches to LLM serving evaluation trade off accuracy, cost, and coverage, and each models a subset of the serving pipeline, which bounds the dynamics (\S\ref{sec:coupling}) it can reproduce. ``Profiling required'' is the calibration granularity: ``per config'' a run per (model, GPU, TP), ``per GPU'' a run per GPU type and precision, ``once'' a single global fit for all configurations, and ``none'' no fit. BLIS alone pairs a latency model needing only a one-time global fit with a coupled full-pipeline simulation and composable policies.}
  \label{tab:landscape}
  \footnotesize
  \setlength{\tabcolsep}{2pt}
  \begin{tabular}{@{}lccccc@{}}
    \toprule
     & \shortstack{LLM-Optimizer\\\cite{bentoml2024llmoptimizer}} & \shortstack{Vidur\\\cite{agrawal2024vidur}} & \shortstack{AIConfigurator\\\cite{xu2026aiconfigurator}} & \shortstack{LLMServingSim\\\cite{cho2026llmservingsim2}} & BLIS \\
    \midrule
    Latency model          & analytical   & profiled       & profiled     & profiled       & \shortstack{analytical\\+ learned} \\
    Tail latency (P90/P99) & no           & yes            & no           & yes            & yes \\
    Profiling required     & none         & per config     & per GPU      & per config     & once \\
    Composable policies    & none         & limited        & none         & limited        & full \\
    \bottomrule
  \end{tabular}
\end{table}

\subsection{Design Requirements}
\label{sec:requirements}

The dynamics of \S\ref{sec:coupling} and the gaps in \S\ref{sec:existing} together impose four requirements on a simulator intended to support configuration search and policy evolution.

\begin{enumerate}
  \item \textbf{End-to-end coupled simulation (R1).} The simulator must model the full pipeline as interacting components rather than isolated stages, to represent emergent effects such as preemption cascades and routing-induced overload (\S\ref{sec:coupling}).
  \item \textbf{Generalization without per-configuration profiling (R2).} It must predict latency for new (model, GPU, TP) combinations from architectural parameters, without re-running calibration for every configuration.
  \item \textbf{Composable policy interfaces (R3).} Each policy axis (admission, routing, scheduling) must be replaceable independently and without modifying the simulator core, so that both manual sweeps and automated search can substitute candidate policies.
  \item \textbf{Fast, deterministic execution (R4).} A run must advance through simulated events rather than wall-clock time, so that hours of serving replay in seconds on a CPU, and must produce identical output for a given seed, so that the simulator can serve as a stable fitness function for iterative optimization.
\end{enumerate}

\noindent Section~\ref{sec:design} addresses each requirement: R1 through the coupled simulation pipeline, R2 through the latency model, R3 through pluggable interfaces, and R4 through CPU-only deterministic execution.

%% file: design.tex

\section{System Design}
\label{sec:design}

BLIS is organized so that each of the four requirements of \S\ref{sec:requirements} is met by a distinct part of the design: the architecture and execution core (\S\ref{subsec:overview}, R4) grounds the rest; the latency model (\S\ref{subsec:latency}, R2) is the central novelty, and \S\ref{subsec:fitting} describes how its shared coefficients are fit once and why the pairing of this model with simulation is the source of BLIS's accuracy; the simulation pipeline (\S\ref{subsec:pipeline}, R1) couples the components end to end; and the policy interfaces (\S\ref{subsec:policy}, R3) make admission, routing, and scheduling independently replaceable.

\subsection{Overview}
\label{subsec:overview}

\begin{figure*}[tbp]
  \centering
  \includegraphics[width=0.8\fulltextwidth]{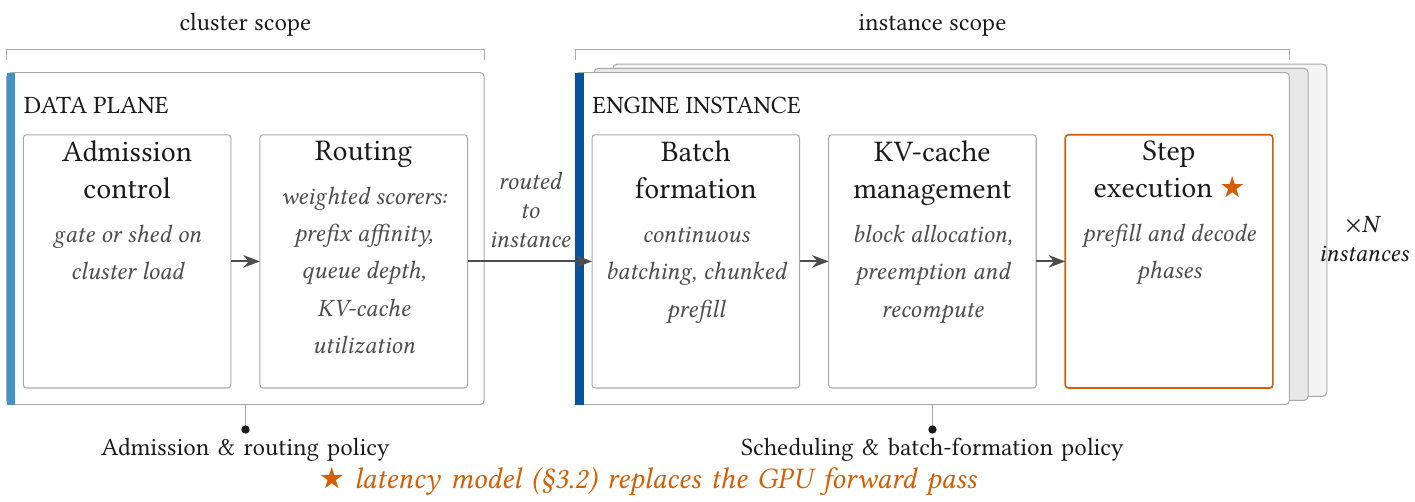}
  \caption{Layered architecture of BLIS. Each layer exposes a single policy interface; the engine layer replaces GPU step execution with the latency model and is otherwise structurally identical to a real instance.}
  \label{fig:arch}
\end{figure*}

BLIS models an LLM serving deployment as two layers that mirror the structure of a production stack (\S\ref{sec:pipeline}).
The \emph{engine layer} executes batches on a single instance: it forms batches, manages the KV cache, and advances token generation.
The \emph{data plane} admits requests and routes them to instances.
Each layer exposes the serving behavior it governs through a policy interface (\S\ref{subsec:policy}), so a deployment is described by the policies plugged into each layer, not by changes to the simulator.
The implementation and experiments are grounded at the engine layer in vLLM~\cite{kwon2023vllm}, the most widely adopted open-source single-instance LLM serving engine, and at the data plane in llm-d~\cite{llmd2024}, an open-source project for distributed LLM serving built on vLLM; both are representative of production deployments.
Figure~\ref{fig:arch} shows the two layers and the policy interface exposed at each.
The only departure from a real deployment is the engine's execution step: where a production instance runs the forward pass on a GPU, BLIS substitutes the latency model of \S\ref{subsec:latency}, otherwise reproducing the same control flow.

\smallskip\noindent\textbf{Discrete-event execution.}
BLIS advances by event rather than by fixed time step.
The simulator maintains a single timeline of scheduled events (request arrival, step completion, snapshot refresh, preemption) and processes them in timestamp order, advancing the simulated clock directly to the next event.
A \emph{snapshot refresh} republishes each instance's load signals (queue depth, KV-cache utilization, batch size) to the routing and admission layer, which act on these periodically refreshed snapshots rather than live instance state (\S\ref{subsec:pipeline}).
A run therefore costs work proportional to the number of events rather than the wall-clock duration simulated, replaying hours of serving in seconds on a CPU (R4).
This is the standard discrete-event simulation (DES) method~\cite{banks2010des}; our contribution is applying it to the coupled serving dynamics of \S\ref{sec:coupling} that a steady-state model misses.
Algorithm~\ref{alg:des} summarizes the loop and shows where each layer of the architecture acts on an event.

\begin{algorithm}[tbp]
\caption{BLIS discrete-event simulation loop.}
\label{alg:des}
\begin{algorithmic}[1]
\State Initialize event queue $Q$ as a min-heap on $(t,\text{priority},\text{seq})$
\State Seed each subsystem's RNG; enqueue request arrivals
\While{$Q$ is not empty}
  \State $e \gets$ pop earliest event from $Q$; advance clock to $e.t$
  \If{$e$ is an arrival}
    \State \textbf{data plane:} admission, then routing to an instance
  \ElsIf{$e$ is a step completion}
    \State \textbf{engine:} form batch, manage KV cache, call latency model (\S\ref{subsec:latency})
    \State enqueue the next step if work remains \Comment{work-conserving}
  \ElsIf{$e$ is a snapshot refresh}
    \State publish instance state to the data plane \Comment{stale-signal feedback}
  \ElsIf{$e$ is a preemption}
    \State free KV blocks and mark victims for recomputation
  \EndIf
\EndWhile
\end{algorithmic}
\end{algorithm}

\smallskip\noindent\textbf{Separating mechanism from policy.}
The code is split into two layers that correspond to this distinction.
A \emph{simulation kernel} provides the discrete-event machinery: the event queue, the simulated clock, partitioned random-number generation, and statistics collection.
It is agnostic to LLM-serving specifics, ordering and dispatching events unchanged across GPUs, models, and configurations.
A set of \emph{domain modules} (router, scheduler, KV-cache manager, latency model, batch-formation logic) encodes all inference-serving behavior.
These modules hold the domain knowledge while consuming and producing events, so a new policy (\S\ref{subsec:policy}) is introduced by implementing one module without touching the kernel.

\smallskip\noindent\textbf{Determinism.}
The event queue is a min-heap keyed on (timestamp, priority, sequence identifier), with cluster-level ties broken by lowest instance index, and randomness is partitioned per subsystem so no stochastic component perturbs another's event order.
A given seed thus produces byte-identical output across runs (R4), as manual sweeps and automated search require to use the simulator as a fitness function (\S\ref{sec:evolution}).

\subsection{Latency Model}
\label{subsec:latency}

The latency model predicts the execution time of one batch step and is what lets BLIS generalize to unprofiled (model, GPU, TP) configurations (R2).
Prior simulators obtain step time by profiling, requiring a calibration run before predicting on a new model, GPU, or parallelism degree (\S\ref{sec:existing}).
BLIS instead expresses step time as a closed-form function of model architecture (hidden dimension, layer and head counts, intermediate dimension, expert count, quantization precision), hardware specifications (peak FLOP/s, HBM bandwidth), and parallelism degree, all read from a model's published \texttt{config.json} and the GPU's datasheet.

\smallskip\noindent\textbf{Model structure.}
The latency model is built in two layers.
The first computes closed-form roofline bounds~\cite{williams2009roofline} at the granularity of a whole forward-pass step, as in Splitwise~\cite{patel2024splitwise} and DistServe~\cite{zhong2024distserve}: a compute bound divides an operation's floating-point operations $F$ by the GPU's peak throughput $P$, and a memory bound divides the bytes moved $M$ by the relevant bandwidth, HBM bandwidth $W$ on-device and interconnect bandwidth $W_{\text{net}}$ across shards.
Each basis function therefore carries units of time: a compute term divides operations by operations per second, and a memory term divides bytes by bytes per second.
Six basis functions cover the dominant costs of a step, grouped by phase: prefill (pf), decode (dc), weight loading, and tensor-parallel (tp) communication:
\begin{equation}
\label{eq:basis}
\begin{aligned}
T^{\text{pf}}_{\text{comp}} &= F^{\text{pf}}/P, &\quad T^{\text{pf}}_{\text{kv}} &= M^{\text{pf}}_{\text{kv}}/W, &\quad T_{\text{weight}} &= M_{\text{w}}/W, \\
T^{\text{dc}}_{\text{comp}} &= F^{\text{dc}}/P, &\quad T^{\text{dc}}_{\text{kv}} &= M^{\text{dc}}_{\text{kv}}/W, &\quad T_{\text{tp}} &= M_{\text{tp}}/W_{\text{net}}.
\end{aligned}
\end{equation}
The two prefill terms ($T^{\text{pf}}_{\text{comp}}$, $T^{\text{pf}}_{\text{kv}}$) cover prompt processing, including the KV-cache write; the two decode terms ($T^{\text{dc}}_{\text{comp}}$, $T^{\text{dc}}_{\text{kv}}$) cover token generation, where reading the accumulated context dominates bandwidth.
Weight loading ($T_{\text{weight}}$) is incurred once per step, since a single forward pass serves the whole batch~\cite{kwon2023vllm}, and $T_{\text{tp}}$ is the all-reduce across shards.
Attention compute follows FlashAttention's IO-aware accounting~\cite{dao2022flashattention}, where new tokens attend within on-chip SRAM so that only past context contributes HBM traffic; quantization and grouped-query attention enter through the same parameters, the effective bytes per parameter and the key-value head count.

The second layer closes the gap between these peak bounds and observed execution time, where real kernels run below peak through memory access patterns, launch overhead, synchronization, and scheduling.
Scaling the roofline by fixed per-phase utilization factors, the conventional analytical approach, cannot track how this gap shifts with batch composition and load; BLIS therefore learns the scaling from end-to-end traces rather than fixing it, a choice we isolate empirically in \S\ref{subsec:ablations}.
It combines the basis functions with learned scaling coefficients and additive overhead terms that capture costs the roofline omits:
\begin{equation}
\label{eq:latency}
\begin{split}
T_{\text{step}} =\ & \beta_{1a} T^{\text{pf}}_{\text{comp}} + \beta_{1b} T^{\text{pf}}_{\text{kv}} + \beta_{2a} T^{\text{dc}}_{\text{comp}} + \beta_{2b} T^{\text{dc}}_{\text{kv}} \\
& + \beta_3 T_{\text{weight}} + \beta_4 T_{\text{tp}} + \beta_5 L + \beta_6 B + \beta_7 + \beta_8 n_{\text{MoE}},
\end{split}
\end{equation}
where $L$ is the number of layers, $B$ the batch size, and $n_{\text{MoE}}$ the number of interleaved mixture-of-experts (MoE) layers.
The first four coefficients ($\beta_{1a}$--$\beta_{2b}$) are dimensionless scaling factors that absorb the gap between peak and observed throughput, fit once across all configurations rather than set per GPU.
The remaining terms are fixed overheads the roofline does not model: a per-layer cost ($\beta_5$, kernel launches and normalization), a per-request cost ($\beta_6$, scheduling and mask construction), a per-step constant ($\beta_7$, synchronization and sampling), and a per-MoE-layer cost ($\beta_8$).
Because $L$, $B$, and $n_{\text{MoE}}$ are dimensionless counts, $\beta_5$--$\beta_8$ carry units of time, so every term in Equation~\ref{eq:latency} contributes a time.
Three further coefficients ($\alpha_0$--$\alpha_2$) account for per-request CPU and framework overhead: request ingestion ($\alpha_0$), detokenization ($\alpha_1$), and per-token streaming ($\alpha_2$).
All coefficients are shared and fit once jointly across diverse (model, GPU, TP) configurations by Bayesian optimization from end-to-end latency traces, with no operator-level profiling (Appendix~\ref{app:training}).

The decomposition in Equation~\ref{eq:latency} doubles as a check on whether the fit reflects the underlying behavior or merely absorbs error.
Each phase carries both a compute and a memory term, but by the roofline only one should bind: prefill is compute-bound and decode memory-bound, so the prefill-memory ($\beta_{1b}$) and decode-compute ($\beta_{2a}$) terms should carry no weight.
The fit confirms this, leaving the prefill-compute and decode-memory factors as the only active terms; it thus recovers the roofline regime rather than spreading weight across all terms, indicating that the physics prior carries the prediction, and the overhead terms take interpretable values (Appendix~\ref{app:training}).

\smallskip\noindent\textbf{Generalization.}
Because the basis functions depend only on published parameters and the coefficients are shared, an unprofiled model, GPU, or TP requires no calibration run: BLIS substitutes the new parameters into the same equations and applies the existing coefficients.
We use the full form by default; \S\ref{sec:eval} evaluates its accuracy across unprofiled models, GPUs, and parallelism degrees.

\subsection{Fitting the Model}
\label{subsec:fitting}

BLIS's accuracy comes from pairing the latency model of \S\ref{subsec:latency}, whose coefficients are fit once across real-system experiments, with the full-pipeline simulation of \S\ref{subsec:overview}, which advances the coupled dynamics of \S\ref{sec:coupling} rather than a steady-state average.
Existing tools supply one element but not both (\S\ref{sec:existing}). Closed-form estimators predict from specifications but do not simulate load-dependent dynamics, while profiled-operator simulators reproduce those dynamics but require a calibration run per new model, GPU, or configuration.
This subsection describes how the shared coefficients are obtained.

\smallskip\noindent\textbf{Training data.}
The coefficients are fit on 15 ground-truth experiments collected from vLLM deployments on H100 GPUs, each recording per-request TTFT, end-to-end latency, inter-token latency, and throughput under a multi-stage workload profile with controlled arrival rates and token-length distributions.
The experiments span six model families (Llama-2-7B, Llama-3.1-70B, Llama-4-Scout-17B-16E, Mistral-Nemo-12B, Qwen2.5-7B, Yi-34B), three architectural classes (dense, grouped-query attention, and interleaved mixture-of-experts with FP8 quantization), and tensor-parallelism degrees of 1, 2, and 4.
This diversity forces the coefficients to generalize: a single fit must hold for dense models where the MoE terms vanish, MoE models where the interleave overhead dominates, single-GPU deployments where the TP term is zero, and multi-GPU deployments where communication cost is significant.
No evaluation model (\S\ref{sec:eval}) appears in the fitting set by design. The evaluation models differ in parameter count from their training counterparts (for example, Llama-2-70B against the training set's Llama-2-7B), and three families (Mixtral, CodeLlama, Qwen3) are absent from training entirely, while two GPU types (A100, L40S) are used only in evaluation.
The accuracy reported in \S\ref{sec:eval} therefore reflects out-of-distribution generalization rather than interpolation within the training regime (\S\ref{sec:eval} enumerates the evaluation models and Appendix~\ref{app:coverage} the full configuration set).

\smallskip\noindent\textbf{Two-loop procedure.}
The fit separates \emph{what the model computes} (its structure) from \emph{how well it fits} (its coefficients) through two nested loops, summarized in Algorithm~\ref{alg:training}.
The outer loop evolves the model structure: examining per-experiment residuals from the previous iteration, identifying a systematic error (for example, MoE underprediction), and hypothesizing a structural change (adding the MoE overhead term $\beta_8$, correcting a basis function, or splitting a combined term).
The inner loop holds the structure fixed and searches for the coefficients that minimize the loss
\begin{equation}
\label{eq:training-loss}
\mathcal{L} = \text{RMSE}\bigl[\text{APE}(\text{TTFT}_i)\bigr] + \text{RMSE}\bigl[\text{APE}(\text{E2E}_i)\bigr],
\end{equation}
where $\text{TTFT}_i$ and $\text{E2E}_i$ are the mean TTFT and end-to-end latency for experiment $i$ and the RMSE is taken over the 15 per-experiment absolute-percentage-error (APE) values.
Taking the RMSE across experiments penalizes variance: a model accurate on one architecture but not another incurs high loss even when its mean error is low, so the fit is pushed toward coefficients that hold across the training diversity rather than favoring any one configuration.
The inner loop selects its optimizer by search-space dimensionality and iteration maturity.
Early iterations use TPE Bayesian optimization~\cite{bergstra2011tpe} via Optuna~\cite{akiba2019optuna} for high-dimensional exploration (up to 2{,}000 trials over all coefficients jointly); as the structure stabilizes and the search narrows, it shifts to CMA-ES~\cite{hansen2003cmaes} for joint multi-parameter refinement and golden-section search~\cite{kiefer1953golden} for one-dimensional polish of individual coefficients, and late iterations combine the three in a pipeline (CMA-ES for global exploration, TPE warm-started with prior trials, golden section for final convergence).
In all cases, coefficients are bounded to physically motivated ranges and warm-started from the previous outer iteration.
Separating the loops keeps structural changes motivated by physics rather than by overfitting: the inner loop cannot mask a missing term by distorting other coefficients, because the per-experiment residuals expose where the model form itself is inadequate.

\begin{algorithm}[tbp]
\caption{Two-loop latency-model fitting.}
\label{alg:training}
\begin{algorithmic}[1]
\State \textbf{Input:} initial model structure $f_0$, training experiments $\{E_1, \ldots, E_{15}\}$
\State \textbf{Output:} final structure $f^*$ and coefficients $(\alpha^*, \beta^*)$
\Statex
\For{each outer iteration $k = 1, 2, \ldots$} \Comment{Structural evolution}
  \State Examine per-experiment errors from iteration $k{-}1$
  \State Identify systematic residuals (e.g., MoE underprediction)
  \State Hypothesize structural change $\rightarrow$ new model form $f_k$
  \State Define coefficient search bounds $\mathcal{B}_k$
  \Statex
  \State Select optimization strategy \Comment{Inner loop}
  \For{each candidate $(\alpha, \beta) \in \mathcal{B}_k$ proposed by optimizer}
    \State Instantiate the latency model with $f_k$ and $(\alpha, \beta)$
    \State Run BLIS on all 15 experiments in parallel
    \State Compute $\mathcal{L} = \text{RMSE}[\text{APE}(\text{TTFT}_i)] + \text{RMSE}[\text{APE}(\text{E2E}_i)]$
  \EndFor
  \State $(\alpha_k^*, \beta_k^*) \gets \arg\min \mathcal{L}$
  \State \textbf{if} loss converged \textbf{then break}
\EndFor
\State \Return $(f_k, \alpha_k^*, \beta_k^*)$
\end{algorithmic}
\end{algorithm}

\smallskip\noindent\textbf{What the fit recovers.}
The converged coefficients are physically interpretable rather than opaque.
As \S\ref{subsec:latency} notes, the prefill-memory ($\beta_{1b}$) and decode-compute ($\beta_{2a}$) scaling factors fit to zero, so the fit recovers the roofline regime (compute-bound prefill, memory-bound decode) rather than spreading weight across redundant terms.
The additive overheads take values consistent with their physical sources on H100: tens of microseconds per layer ($\beta_5$, kernel launches and normalization), a few microseconds per request ($\beta_6$, scheduling and mask construction), on the order of a hundred microseconds per step ($\beta_7$, synchronization and sampling), and hundreds of microseconds per interleaved MoE layer ($\beta_8$).
That these independently fit values land in plausible ranges indicates the physics prior, not the free parameters, carries the prediction (full coefficient roles in Appendix~\ref{app:training}).

\smallskip\noindent\textbf{Why the fit transfers without retraining.}
The coefficients are shared across all configurations, and the only architecture-dependent component is the basis functions (Equation~\ref{eq:basis}), computed from the published HuggingFace \texttt{config.json} and the GPU datasheet.
An unprofiled model, GPU, or TP degree is thus handled by evaluating the same basis functions on its published parameters, with no re-fitting; a genuinely new architecture class (for example, sliding-window or multi-head latent attention) needs only an added basis term and a re-run of the inner loop, leaving existing coefficients valid.
The fit therefore requires only public inputs, with no operator-level profiling, kernel tracing, or model weights (Appendix~\ref{app:training}).

\subsection{Simulation Pipeline}
\label{subsec:pipeline}

The pipeline carries a request from arrival to completion through the layers of \S\ref{subsec:overview}, and its coupling is what makes BLIS an end-to-end simulation rather than a sequence of independent stages (R1).
Figure~\ref{fig:eventflow} (Appendix~\ref{app:lifecycle}) traces this lifecycle and marks the snapshot feedback path from instance state back to the routing and admission layer that drives the feedback loops of \S\ref{sec:coupling}.

\smallskip\noindent\textbf{Engine layer.}
At an instance, an admitted request enters a wait queue ordered by the scheduling policy (first-come-first-served, priority, or shortest-job-first).
Batch formation then admits requests from the wait queue up to the configured limits on running requests and scheduled tokens, allocating KV-cache blocks for each.
When the KV cache cannot satisfy a new allocation, the policy preempts running requests to free blocks, selecting victims by arrival order or priority. A preempted request must later recompute its evicted KV state before resuming, exactly the cascade described in \S\ref{sec:coupling}.
Step execution calls the latency model of \S\ref{subsec:latency} on the formed batch, advances the clock by the predicted step time, and schedules the next step.
Long prompts are split across steps by chunked prefill ~\cite{agrawal2024sarathi}, interleaving a large prefill with ongoing decodes rather than stalling them in one step.
The engine maintains a work-conserving invariant: if the wait queue is non-empty after a step completes, a subsequent step is always scheduled, so the instance never idles while work is pending.

\smallskip\noindent\textbf{Data plane.}
At cluster scale, a request first passes admission control, which may gate or shed it based on the current router view of cluster load. Admission policies range from admitting all traffic to saturation-based flow control.
An admitted request is then routed by a scorer framework that combines weighted signals (queue depth, KV-cache utilization, prefix affinity, and load balance) into a per-instance score, the same composition used by the production llm-d router~\cite{llmd2024}.
These decisions read periodically refreshed snapshots of instance state (queue depth, KV utilization, batch size, in-flight count) rather than instantaneous truth, reproducing the staleness-induced feedback of \S\ref{sec:coupling}.
A cluster-level event queue orchestrates the per-instance queues so that this coupling across instances is simulated rather than approximated.

\smallskip\noindent\textbf{Workload generation.}
The pipeline is driven by one of three workload sources.
Synthetic workloads draw arrivals from a configurable process (Poisson, gamma, Weibull, or constant) with prompt- and output-length distributions and per-request SLO tiers (critical, standard, sheddable).
Trace replay reissues captured request traces with either their original timing or simulation-computed timing.
Closed-loop session workloads model multi-turn conversations in which each follow-up arrives a think-time after the previous response completes.

\subsection{Pluggable Policy Interfaces}
\label{subsec:policy}

The composability requirement (R3) is met by exposing each policy axis as a single interface that the simulation core calls but does not implement.
Admission decides whether to admit or shed a request; routing selects a target instance through the weighted scorer composition above; the instance scheduler orders the wait queue before batch formation; and batch formation governs the preemption strategy and batch composition.
A new policy along any axis is introduced by implementing the corresponding interface, without changing the kernel or the other modules, so both manual sweeps and automated search can substitute candidates freely.
The routing interface mirrors the llm-d production abstraction directly: a deployment composes named, weighted scorers (for example \texttt{prefix-affinity:2}, \texttt{queue-depth:1}, \texttt{kv-utilization:1}) exactly as in the real system~\cite{llmd2024}, so a policy studied in simulation transfers to production without reimplementation.
Determinism (R4) makes comparing two policies a controlled experiment: with the same workload and seed, the only difference between runs is the policy under test. This interface structure supports both the configuration sweeps of \S\ref{sec:config} and the policy evolution of \S\ref{sec:evolution}.

%% file: experiments_accuracy.tex
\section{Accuracy Study}
\label{sec:eval}

We aim to answer three questions:
(1) Does BLIS generalize across models, hardware, and workloads without per-model or per-GPU profiling runs?
(2) Where does it sit on the accuracy-speed tradeoff relative to existing tools?
(3) Under what conditions does it break, and why?
We describe the setup (\S\ref{subsec:setup}), compare tools head-to-head (\S\ref{subsec:headtohead}), analyze the accuracy-speed tradeoff (\S\ref{subsec:costaccuracy}), ablate BLIS's modeling choices (\S\ref{subsec:ablations}), and characterize failure modes (\S\ref{subsec:failure-modes}).

\subsection{Experimental Setup}
\label{subsec:setup}

\smallskip\noindent\textbf{Models.}
We select six models spanning dense and MoE architectures, 8B--141B parameters, and distinct attention mechanisms:
four dense models (Llama-3.1-8B, Qwen3-14B, CodeLlama-34B, and Llama-2-70B) and two MoE models (Mixtral-8x7B, 47B total/13B active; Mixtral-8x22B, 141B total/39B active).
None of these six models appears in the training set used to fit BLIS's coefficients; the two sets are disjoint by design (\S\ref{subsec:fitting}), so the evaluation tests generalization to unseen parameter counts and model families within the dense (multi-head and grouped-query attention) and mixture-of-experts transformer classes BLIS covers (\S\ref{subsec:latency}).

\smallskip\noindent\textbf{Hardware.}
Experiments run on three NVIDIA GPU types spanning distinct memory-bandwidth, compute, and interconnect profiles: H100-80GB SXM (NVLink), A100-80GB SXM (NVLink), and L40S (48\,GB, PCIe), testing whether the latency model generalizes across hardware generations without per-GPU re-profiling.

\smallskip\noindent\textbf{Workloads.}
Each experiment replays a fixed request trace against the real system and all simulators.
Token-length distributions come from ServeGen~\cite{xiang2025servegen}, which regenerates request shapes from production serving traces, and arrival timing from the Azure LLM inference trace~\cite{azure2024publicdataset}. Drawing shapes and arrivals from separate production sources lets each reflect measured behavior, with skewed length distributions that stress scheduling fidelity beyond synthetic workloads.
We instantiate four workload profiles at arrival rates of 4--20\,rps, differing primarily in token-length distribution:
\emph{general-purpose} (input ${\approx}$550, output ${\approx}$250 tokens),
\emph{codegen} (input ${\approx}$570, output ${\approx}$250 tokens),
\emph{roleplay} (input ${\approx}$750, output ${\approx}$250 tokens, longer system prompts), and
\emph{reasoning} (input ${\approx}$1{,}030, output ${\approx}$1{,}450 tokens), which is far more decode-heavy, stressing KV-cache capacity and producing longer queue-drain tails.
All workloads use multi-stage load profiles (stages at different arrival rates) and contain 10.4k requests on average.

\smallskip\noindent\textbf{Configurations.}
vLLM exposes serving parameters controlling the compute/memory tradeoff: chunk size (\texttt{max\_num\_batched\_tokens}) sets how many prefill tokens run per step, and thus how often prefill and decode interleave; GPU memory utilization sets the KV-cache budget; and CPU offloading extends the scheduler's memory hierarchy.
The \emph{baseline configuration} for each model uses the minimum viable tensor-parallel degree, \texttt{max\_num\_batched\_tokens}=2048, \texttt{gpu\_memory\_utilization}=0.9, and CPU offloading disabled; from this baseline we sweep one knob at a time: tensor-parallel degree $\in \{1, 2, 4, 8\}$, \texttt{max\_num\_batched\_tokens} $\in \{1024, 2048, 8192\}$, \texttt{gpu\_memory\_utilization} $\in \{0.9, 0.95\}$, and KV-cache CPU offloading (on/off).

\smallskip\noindent\textbf{Baselines.}
We compare against four tools whose capabilities are summarized in Table~\ref{tab:landscape}; for each, we note only the setup specific to our evaluation.
\begin{itemize}
  \item \emph{Vidur}~\cite{agrawal2024vidur}: we use only its natively shipped profiles (no additional profiling), which exclude MoE architectures and limit hardware and model coverage. Its scheduler follows pre-chunked-prefill vLLM semantics structurally incompatible with modern vLLM~v1, a gap re-profiling alone cannot bridge (Appendix~\ref{app:vidur}).
  \item \emph{LLMServingSim}~\cite{cho2026llmservingsim2}: it ships profiling data only for the RTX PRO 6000, so for our H100, A100-80GB, and L40S targets we profiled all models with its built-in profiler. Its cycle-accurate simulation gives high architectural fidelity at proportionally higher runtime cost, constraining the request volume we can simulate.
  \item \emph{AIConfigurator}~\cite{xu2026aiconfigurator}: we supply model architecture, hardware type, and precision, and read latency from the produced throughput-versus-latency Pareto frontiers.
  \item \emph{LLM-Optimizer}~\cite{bentoml2024llmoptimizer}: evaluates in sub-second time and approximates MoE models as dense.
\end{itemize}
BLIS and Vidur replay full per-request traces.
LLM-Optimizer and AIConfigurator are steady-state estimators that do not natively accept request-rate inputs; we derive their concurrency parameters via throughput-matching (Appendix~\ref{app:concurrency}), whose discretization residual (mean 8.7\% and 12.3\%) is well below their latency prediction error, so the translation is not the source of their inaccuracy.
LLMServingSim runs its cycle-level simulator with the analytical ASTRA-Sim backend~\cite{Won_2023} on a proportionally sampled 1{,}000-request subset per experiment, since the full 10k+ request workloads are infeasible at cycle granularity, and is compared against vLLM ground truth on that subset.

\smallskip\noindent\textbf{Ground truth and metrics.}
Ground-truth measurements are collected using inference-perf~\cite{inferenceperf2024} on vLLM v0.15.1~\cite{kwon2023vllm} (which runs the V1 engine architecture) serving deployments across all three GPU types.
All experiments completed successfully with zero failed requests.
To characterize the real system's intrinsic noise floor, we repeated a subset of ground-truth experiments three times (independent replicas on the same cluster, spanning 4--40\,rps across three workload profiles).
The coefficient of variation (CV) across replicas is below 5\% for median E2E and below 6\% for E2E P90 at moderate arrival rates (4--15\,rps), and median ITL remains stable across all rates (median CV ${<}$4\%); CV rises at higher arrival rates (16--40\,rps) where queueing effects amplify scheduling-order perturbations.
A single ground-truth measurement is therefore a stable reference for the operating points where the accuracy study concentrates.
We record per-request time-to-first-token (TTFT), inter-token latency (ITL), and end-to-end (E2E) latency.
The primary accuracy metric is mean absolute percentage error (MAPE):
\begin{equation}
  \text{MAPE} = \frac{1}{N}\sum_{i=1}^{N} \left| \frac{\hat{y}_i - y_i}{y_i} \right| \times 100\%
\end{equation}
where $\hat{y}_i$ is the simulator's prediction and $y_i$ is the measured value for experiment $i$.
We report E2E MAPE as the primary metric and include TTFT and ITL breakdowns where they reveal distinct behavior.
The evaluation comprises 36 experiments in total, enumerated in Appendix~\ref{app:coverage}; across them, BLIS achieves 6.7\% median E2E MAPE, and predicts E2E P90 and P99 latency with 7.5\% and 9.2\% median error (\S\ref{subsec:tail}).

\subsection{Head-to-Head Comparison}
\label{subsec:headtohead}

We compare all tools across the full experiment set, slicing results along four axes: model architecture, GPU type, workload type, and configuration knobs.
Each tool predicts accurately only within the regime it models, so we report prediction error where a tool is in-regime and identify configurations that fall outside it rather than counting them as latency-model error.

\begin{figure}[tbp]
  \centering
  \includegraphics[width=\columnwidth]{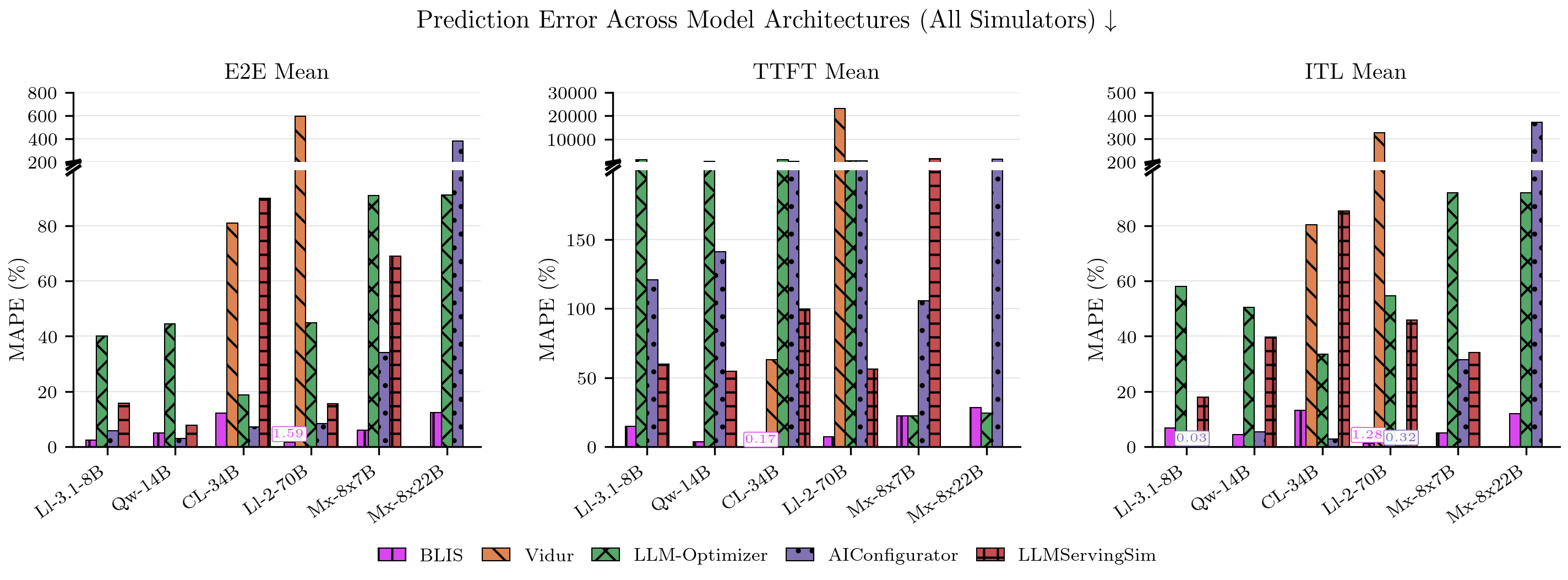}
  \caption{Per-tool MAPE across six models on H100 (E2E, TTFT, ITL). Vidur's shipped profiles exclude MoE; its TTFT overprediction reflects a structural scheduler mismatch with vLLM~v1 (Appendix~\ref{app:vidur}), not latency-model error.}
  \label{fig:accuracy-models}
\end{figure}

\smallskip\noindent\textbf{Accuracy across models.}
Figure~\ref{fig:accuracy-models} reports per-tool MAPE across six models on H100 under baseline parameters (\S\ref{subsec:setup}).
BLIS shows the lowest E2E error on four of the six models and the lowest TTFT error on five of the six.
AIConfigurator is moderately accurate where it runs and strongest on ITL, where its analytical time-per-output-token model matches the decode-bound regime and edges BLIS on three models.
Vidur runs only the four dense models; where it does, its TTFT error exceeds 10{,}000\% because it emulates an older vLLM scheduler without chunked prefill, modeling prefill as one blocking step rather than interleaved chunks.
This gap reflects applicability to vLLM~v1 rather than the accuracy of Vidur's latency model, and is structural (Appendix~\ref{app:vidur}).
LLM-Optimizer's E2E error is moderate on dense models (20--50\%) but rises to ${\approx}$90\% on MoE, where its single-request closed-form estimate approximates MoE routing as dense.

\begin{figure}[tbp]
  \centering
  \includegraphics[width=\columnwidth]{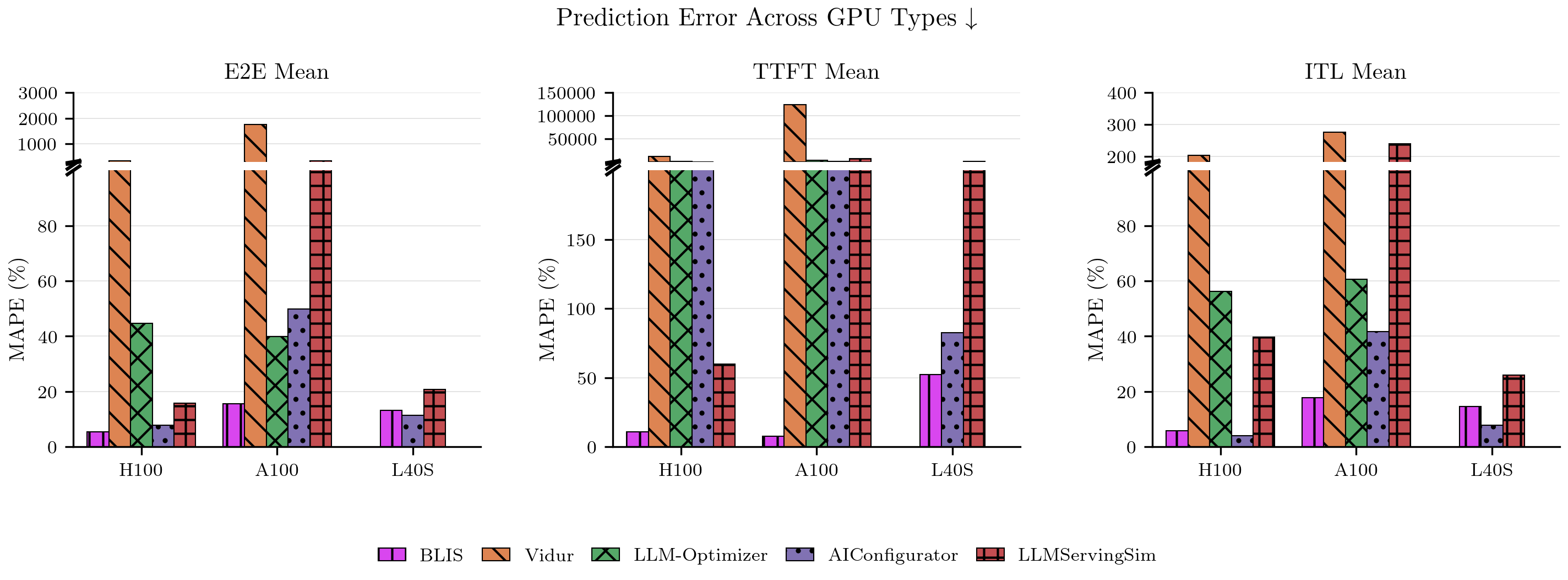}
  \caption{Per-tool MAPE grouped by GPU type (\texttt{mbt}=2048, \texttt{gpu\_mem}=0.9, offloading off; general-purpose workloads). Vidur ships profiles for only a subset of models on A100 and none on L40S.}
  \label{fig:accuracy-hardware}
\end{figure}

\smallskip\noindent\textbf{Accuracy across hardware.}
Figure~\ref{fig:accuracy-hardware} groups experiments by GPU (baseline parameters, general-purpose workloads).
Because BLIS's coefficients are fit only on H100 (Appendix~\ref{app:training}), its A100 and L40S results measure cross-GPU generalization with no per-GPU re-profiling.
Median E2E MAPE is 5.5\% on H100 and 15.7\% and 13.3\% on the unprofiled A100 and L40S, so BLIS predicts GPUs it was never calibrated on from datasheet parameters alone, without a per-GPU profiling run.
BLIS has the lowest E2E error on H100 and A100 and the lowest TTFT error on all three GPUs. On L40S, AIConfigurator edges it on E2E (11.5\% vs 13.3\%) and ITL.
LLM-Optimizer errors are uniformly high because its peak-spec roofline misses per-GPU efficiency differences under batched serving.

\begin{figure}[tbp]
  \centering
  \includegraphics[width=\columnwidth]{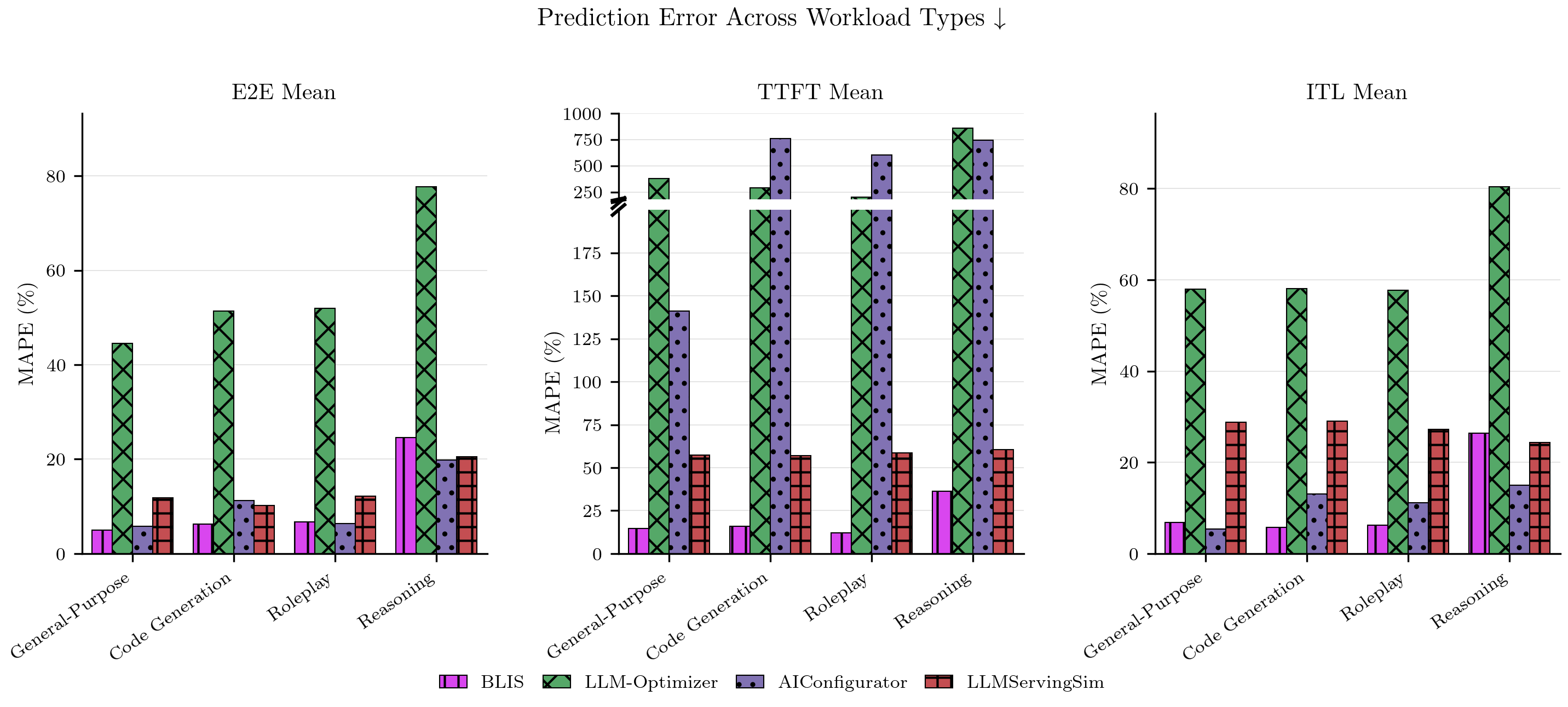}
  \caption{Per-tool MAPE for up to four workload types across Llama-3.1-8B, Qwen3-14B, and Mixtral-8x22B (H100, \texttt{mbt}=2048, \texttt{gpu\_mem}=0.9, offloading off). Vidur excluded (no profiled coefficients for these models).}
  \label{fig:accuracy-workload}
\end{figure}

\smallskip\noindent\textbf{Accuracy across workloads.}
Figure~\ref{fig:accuracy-workload} reports median MAPE for up to four workload types across Llama-3.1-8B, Qwen3-14B, and Mixtral-8x22B (H100, baseline parameters).
BLIS stays below 10\% median E2E MAPE on general-purpose, code generation, and roleplay, rising to 25\% on the decode-heavy reasoning workload, whose longer input and output sequences amplify KV-cache pressure, prefill-chunk scheduling, and queue-drain tails.
LLM-Optimizer and AIConfigurator degrade on code generation and reasoning because their concurrency-based models cannot capture how sequence-length distribution shapes batch composition and step-time variance.

\smallskip\noindent\textbf{Accuracy across configurations.}
Sweeping one knob at a time (TP degree, chunk size, GPU memory utilization, KV-cache offloading) on H100 for a dense model (Llama-3.1-8B) and an MoE model (Mixtral-8x7B), BLIS achieves the lowest E2E MAPE at the baseline and most configurations, whereas LLM-Optimizer and AIConfigurator ignore chunk size, offloading, and memory utilization, predicting identically across configurations.
The full per-knob breakdown appears in Appendix~\ref{app:config-sweep} (Figure~\ref{fig:config-sensitivity}).

\subsection{Accuracy-Speed Tradeoff}
\label{subsec:costaccuracy}

\begin{figure}[tbp]
  \centering
  \includegraphics[width=\columnwidth]{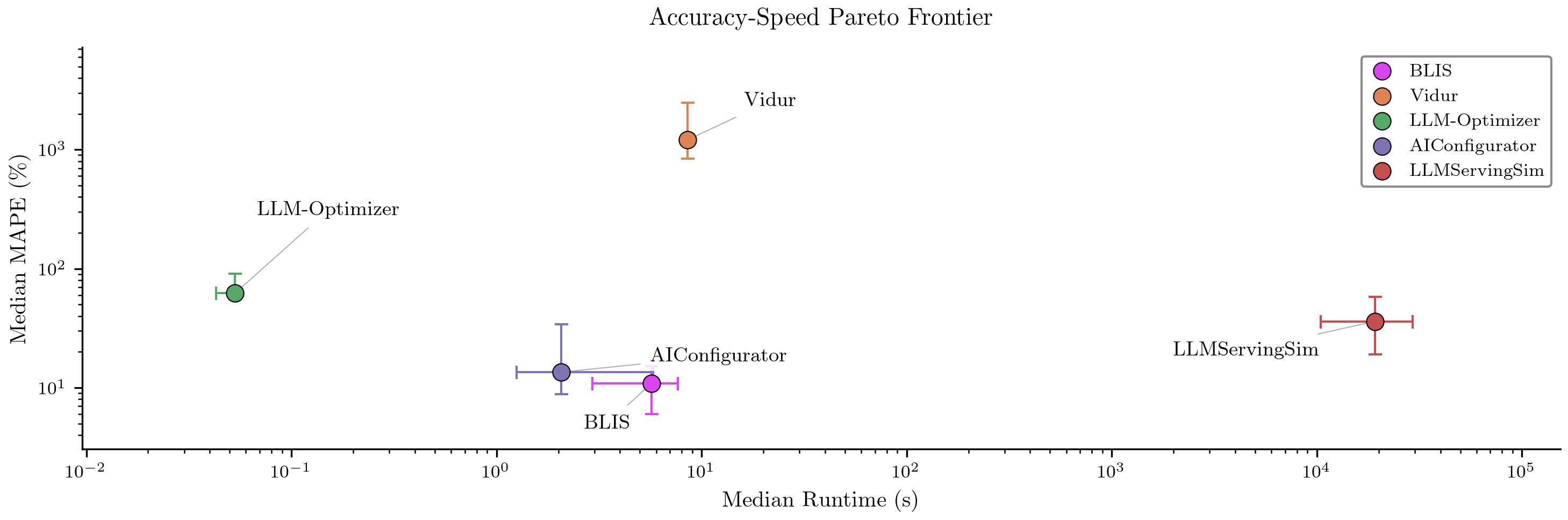}
  \caption{Accuracy versus speed: median runtime (s) versus median MAPE (\%) on log-log axes. Error bars show interquartile range. LLMServingSim is evaluated on a 1{,}000-request subset (\S\ref{subsec:setup}), so its runtime is not directly comparable to the full-trace tools.}
  \label{fig:accuracy-speed}
\end{figure}

Accuracy alone does not determine practical value: per-evaluation speed sets whether exhaustive search is tractable. Figure~\ref{fig:accuracy-speed} plots median MAPE against median wall-clock runtime per simulator on log-log axes across all experiments.
LLM-Optimizer evaluates in 0.1\,s by eliminating queueing simulation, but its analytical estimates cannot capture batch-composition and scheduling effects, at substantially higher error.
BLIS occupies the low-error, moderate-speed region: slower than estimation-based tools but scheduling-aware.
LLMServingSim's cycle-accurate fidelity costs runtime exceeding real execution, and Vidur shows the highest median error among the dense configurations it runs (\S\ref{subsec:headtohead}).
Exact per-tool runtimes and speedups appear in Appendix~\ref{app:runtime} (Table~\ref{tab:accuracy-runtime}).

\subsection{Ablations}
\label{subsec:ablations}

We ablate the learned component of the latency model (\S\ref{subsec:latency}) to quantify the accuracy it adds over a purely analytical baseline.

\smallskip\noindent\textbf{Roofline versus trained-physics.}
The \emph{roofline} configuration uses the same analytical basis functions (Equation~\ref{eq:basis}) but scales them by fixed Model-FLOPs-Utilization (MFU) factors rather than learned coefficients.
These factors are set per phase and GPU from published utilization benchmarks~\cite{shah2024flashattention3, korthikanti2023megatron, reddi2020mlperfinference} (details in Appendix~\ref{app:mfu}).
No overhead terms ($\beta_5$--$\beta_8$, $\alpha_0$--$\alpha_2$) are included; this configuration requires no training.
The \emph{trained-physics} configuration is the full model used by default: it replaces the fixed MFU factors with the learned scaling coefficients ($\beta_{1a}$--$\beta_4$) and additive overhead terms ($\beta_5$--$\beta_8$) of Equation~\ref{eq:latency}, together with the per-request overheads ($\alpha_0$--$\alpha_2$), fit once across a training set of (model, GPU, TP) configurations (Appendix~\ref{app:training}).

\begin{figure}[tbp]
  \centering
  \includegraphics[width=\columnwidth]{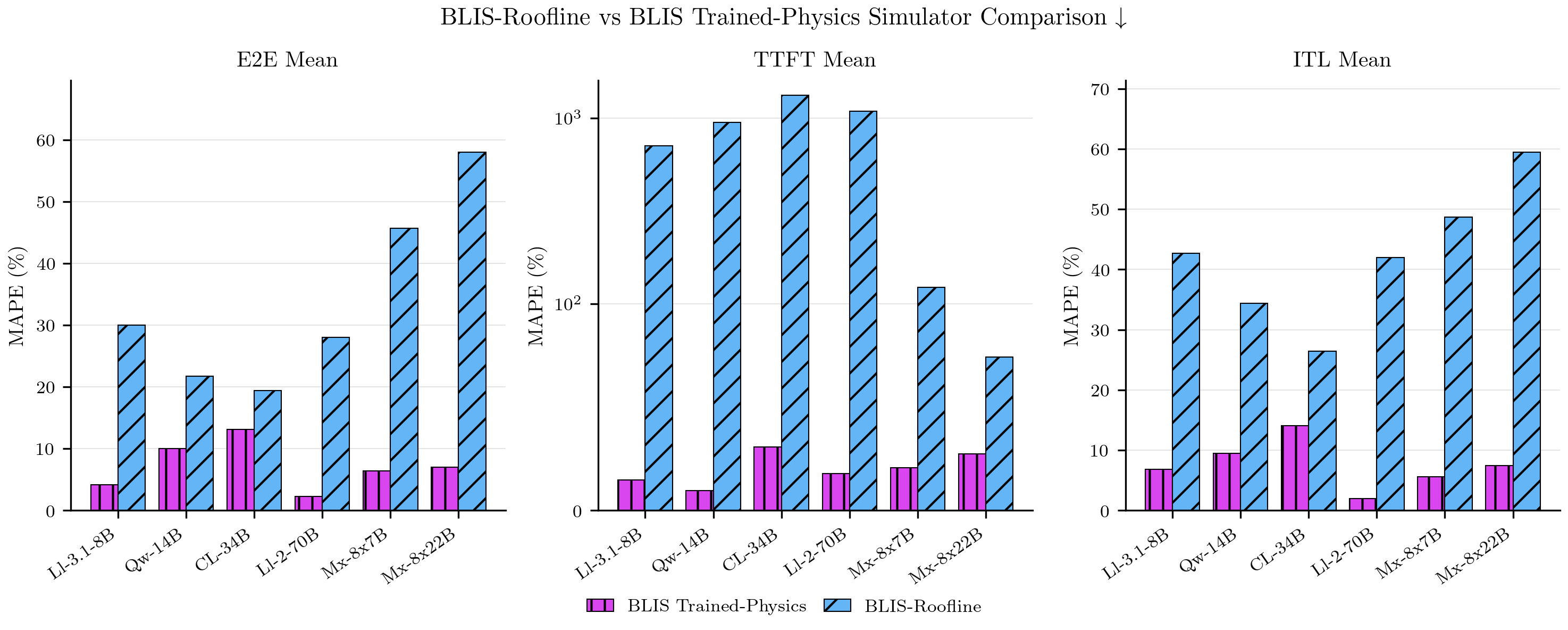}
  \caption{Trained-physics versus roofline configuration per model, showing E2E, TTFT (log scale), and ITL MAPE. The learned scaling and overhead terms reduce error across all models, with the largest gains on TTFT.}
  \label{fig:roofline-vs-trained}
\end{figure}

Figure~\ref{fig:roofline-vs-trained} compares the two across all six models on E2E, TTFT, and ITL.
The roofline configuration systematically under-predicts latency: its fixed MFU factors miss load-dependent overheads from kernel launches, memory bank conflicts, and communication stalls.
The gap is starkest on TTFT (log scale), where roofline incurs $>$100\% MAPE on most models while trained-physics stays substantially lower; it widens on models with more overhead relative to compute, namely CodeLlama-34B and Llama-2-70B (longer memory-bound prefills).
The same coefficients are shared across models without re-fitting, since the captured overhead sources are common across architectures.

\subsection{Tail-Latency Accuracy}
\label{subsec:tail}

By simulating per-request queueing and scheduling, BLIS predicts tail percentiles (P90, P99) that closed-form estimators cannot, and on which the search objectives of \S\ref{sec:config} depend. Across the 36 experiments, BLIS predicts E2E latency to a median MAPE of 6.7\% at the mean, 7.5\% at P90, and 9.2\% at P99; TTFT to 17.6\% at the mean, 21.3\% at P90, and 29.0\% at P99; and ITL to 7.1\% at the mean (inference-perf reports mean ITL only, so ITL has no tail percentiles). Most of the tail error comes from decode-heavy reasoning experiments (\S\ref{subsec:failure-modes}).

\subsection{Failure Modes}
\label{subsec:failure-modes}

We identify two conditions under which BLIS's accuracy degrades.

\smallskip\noindent\textbf{Reasoning workloads.}
The reasoning workload (input ${\approx}$1{,}030, output ${\approx}$1{,}450 tokens) is BLIS's hardest regime (25\% median E2E MAPE): small per-step timing differences compound over ${\sim}$1{,}450 decode iterations, so the simulated scheduling order drifts from reality as long decode sequences hold KV-cache memory and reshape batch-admission timing.

\smallskip\noindent\textbf{All-reduce contention.}
BLIS's analytical all-reduce bandwidth model under-estimates NVLink contention when concurrent batches compete for bandwidth.
On the dense model (Llama-3.1-8B), E2E MAPE rises from 2.5\% at TP=1 to 15.5\% at TP=2 (Appendix~\ref{app:config-sweep}).
The effect is model-dependent rather than monotonic in TP: on Mixtral-8x7B higher parallelism reduces error, while Mixtral-8x22B (TP=8), where all-reduce, expert-dispatch, and memory-footprint errors compound, reaches 12.5\% E2E MAPE on the general-purpose baseline versus 6.0\% for Mixtral-8x7B.

%% file: config_search.tex
\section{Configuration Search}
\label{sec:config}

Deploying an LLM requires jointly choosing tensor-parallelism degree, replica count, batch limits, chunk size, and scheduling, routing, and admission policies: conservative choices over-provision hardware, aggressive ones violate latency SLOs under load.
This space is combinatorially large, and validating each candidate on hardware means provisioning GPUs and running a sustained load test, making exhaustive evaluation prohibitively slow.
BLIS supports tail-latency objectives (P90, P99) and arrival-rate sweeps as first-class search dimensions, enabling the multi-objective exploration shown in Figure~\ref{fig:pareto}.

We compare four tools' ability to search this space for Llama-2-70B on H100 GPUs under a shared-prefix code-generation workload, and validate their top recommendations on real hardware.
We choose Llama-2-70B because all four tools natively support it, aligning the search space as closely as possible across tools.
We use llm-d~\cite{llmd2024} as the deployment platform because it exposes the full parameter surface under test (admission control, routing policy, and vLLM serving configuration) through a unified API.
We evaluate BLIS, Vidur, AIConfigurator, and LLM-Optimizer; LLMServingSim is excluded because it provides no search algorithm and its per-configuration runtime is too slow for practical search (\S\ref{subsec:costaccuracy}).
The four tools differ in how they model latency and what profiling they require, as summarized in Table~\ref{tab:landscape}.

All four tools search TP $\in \{2, 4, 8\}$ and derive replica counts from the GPU budget; Vidur's profiling data does not cover TP=8, restricting its effective search to TP~$\in \{2, 4\}$.
For parameters a tool determines internally (batch sizes, concurrency levels, scheduler caps), we use each tool's own logic rather than imposing external values.
BLIS pairs its simulator with a multi-objective search driven by a declarative YAML specification; an operator targeting a different workload or SLO changes only the spec file.
The search space spans topology, batching (\texttt{max\_num\_seqs}, \texttt{max\_num\_batched\_tokens}), scheduling, routing, admission, and preemption policy (420 configurations, 8 parallel workers).
The search optimizes three objectives (maximize throughput, minimize mean TTFT, minimize GPU count) subject to $\text{TTFT}_\text{mean} \le 300$\,ms and $\text{GPU count} \le 8$.
We use mean TTFT rather than a tail percentile so that all four tools can be compared on the same metric; BLIS and Vidur additionally predict P90/P99, but AIConfigurator and LLM-Optimizer produce only mean estimates.

Of BLIS's 420 configurations, 373 are feasible under the SLO; the Pareto frontier reduces to 2 operating points (TP=4$\times$1 at 4 GPUs, TP=8$\times$1 at 8 GPUs), confirming that topology dominates at moderate utilization.
We deploy each tool's top-2 Pareto configurations (8 total) on H100 GPUs running llm-d under constant 5\,rps arrival (5{,}000 requests, mean input/output 566/247 tokens).

\begin{table}[tbp]
  \centering
  \small
  \caption{Each tool's top-2 recommended configurations with predicted and measured performance. Deployed on H100 GPUs running llm-d under constant 5\,rps arrival (5{,}000 requests). SLO: $\le$300\,ms mean TTFT. Completed denotes the fraction of requests that finished before a 30-minute deadline (arrivals span 1{,}000\,s; a well-provisioned deployment drains all requests well within this window). Note the GPU column: Vidur's low measured TTFT comes at 8 GPUs (\$25.60/hr), double the cost of BLIS's 4-GPU recommendation (\$12.80/hr). Vidur cannot produce a feasible 4-GPU configuration because its scheduler does not model chunked prefill.}
  \label{tab:validation}
  \begin{tabular}{@{}llccrrrc@{}}
    \toprule
    \textbf{Tool} & \textbf{TP$\times$Replicas} & \textbf{GPUs} & \textbf{max\_seqs} & \multicolumn{2}{c}{\textbf{Mean TTFT (ms)}} & \textbf{Completed} & \textbf{SLO} \\
    & & & & Pred. & Meas. & & \\
    \midrule
    BLIS & 4$\times$1 & 4 & 64 & 46 & 96 & 100\% & \checkmark \\
    BLIS & 4$\times$1 & 4 & 256 & 46 & 112 & 100\% & \checkmark \\
    \midrule
    Vidur & 4$\times$2 & 8 & 64 & 188 & 74 & 100\% & \checkmark \\
    Vidur & 4$\times$2 & 8 & 256 & 190 & 124 & 100\% & \checkmark \\
    \midrule
    AICfg & 2$\times$2 & 4 & 20 & 188 & 732 & 100\% & $\times$ \\
    AICfg & 2$\times$3 & 6 & 20 & 188 & 132 & 100\% & \checkmark \\
    \midrule
    LLMOpt & 2$\times$1 & 2 & 4 & 284 & 453{,}677 & 12\% & $\times$ \\
    LLMOpt & 4$\times$1 & 4 & 8 & 284 & 463{,}963 & 48\% & $\times$ \\
    \bottomrule
  \end{tabular}
\end{table}

Table~\ref{tab:validation} reports predicted and measured performance.
The central result is a feasibility separation: BLIS identifies a 4-GPU topology (TP=4$\times$1) that meets the 300\,ms SLO (96\,ms and 112\,ms measured TTFT, 100\% completion), whereas Vidur's pre-chunked-prefill scheduler rejects TP=4$\times$1 and requires 8 GPUs to produce a feasible recommendation (\$25.60/hr versus \$12.80/hr).
Among scheduling-aware tools, BLIS's chunked-prefill model enables a $2\times$ hardware reduction.

AIConfigurator's 4-GPU recommendation (TP=2$\times$2) measures 732\,ms mean TTFT; TP=2 provides insufficient prefill parallelism, saturating at 5\,rps. Its 6-GPU recommendation meets the SLO.
LLM-Optimizer's configurations fail catastrophically (mean TTFT exceeding 450\,s, 12--48\% completion): the tool's own search recommends \texttt{max\_num\_seqs}${\leq}$8, but the workload's steady-state concurrency at 5\,rps is ${\approx}$28 requests in flight, so the low batch limit causes unbounded queue growth.
This failure reflects the tool's single-request model, which does not account for queueing under continuous load, rather than an artifact of our concurrency translation (Appendix~\ref{app:concurrency} derives only the latency-prediction input, not the configuration recommendation).

The critical capability this experiment isolates is distinguishing feasible from infeasible configurations under load, not prediction accuracy on feasible ones.
This experiment uses a single model/GPU/workload combination (Llama-2-70B, H100, code generation) because the accuracy study (\S\ref{sec:eval}) already establishes prediction generalization across six models, three GPU types, and four workloads; the purpose here is to demonstrate that accurate simulation translates into correct feasibility rankings in a search setting, which requires end-to-end hardware validation of recommended configurations rather than additional breadth.
The prediction gap on BLIS's own recommendations does not affect ranking because all TP=4 candidates predict identically, but non-monotonic errors could in principle prevent a Pareto frontier from transferring to hardware.

%% file: case_study.tex
\subsection{Why Policy Evolution Requires BLIS}
\label{subsec:why-evolution}
BLIS models the coupled dynamics of distributed LLM serving and exposes each policy axis (admission, routing, scheduling) as a pluggable interface mirroring a distributed inference stack: a deployment composes named, weighted scorers as in llm-d~\cite{llmd2024}, so a policy studied in simulation transfers smoothly.
This makes BLIS a suitable inner loop for a policy-evolution workflow in which human operators or AI agents explore, evaluate, and refine serving policies.

Existing tools (\S\ref{sec:existing}) either omit admission and routing or do not expose them as replaceable interfaces, so introducing a new policy requires modifying simulation internals.

\smallskip\noindent\textbf{The admission-control problem.}
\label{subsec:admission-problem}
An admission controller decides, per request, whether to admit or shed it.
The default policy admits all requests until the cluster saturation signal reaches a threshold, then sheds all sheddable-tier requests outright.
Per-instance saturation $s_i = \max(\tfrac{\text{queue\_depth}}{\text{max\_queue}}, \tfrac{\text{kv\_used}}{\text{kv\_total}})$ is the larger of two normalized utilizations (queue depth and KV-cache occupancy); the cluster-level signal is the average $s = \frac{1}{N}\sum_i s_i$ across $N$ instances.

This binary threshold fires only at deep queues and near-full KV caches, so critical-tier requests admitted just before it suffer elevated TTFT from contention with queued sheddable traffic (\S\ref{sec:coupling}).
A policy shedding sheddable traffic gradually, before the threshold, protects critical-tier latency.
The design space spans the shedding-curve shape, the saturation definition, and threshold parameters.

\smallskip\noindent\textbf{Simulation-driven policy evolution.}
\label{subsec:policy-search}
Evolving a better policy on real hardware is impractical: a single workload evaluation on a GPU cluster takes 17--42 minutes (measured on H100 nodes running Qwen3-14B with 10{,}000 requests), and evaluating many candidate policies across multiple workloads multiplies this cost.
In BLIS the same evaluation completes in seconds.

We drove the evolution with Nous~\cite{nous2025}, an open-source agentic framework that proposes candidate implementations, measures their effect, and refines the next candidate from the outcome. Each candidate admission policy and workload was evaluated with BLIS as the evolution backend.
The workflow applies equally to a human expert; we use an AI agent to exercise the loop.

BLIS's determinism (\S\ref{subsec:policy}, R4) makes each run a controlled experiment: with the same workload and seed, the only difference between two runs is the policy under test.
The evolution exploited this by testing each candidate across multiple seeds and workloads, catching failure modes (for example, excessive shedding at low load) that a single-workload evaluation would miss.

The objective was to reduce critical-tier latency relative to the default policy by at least 30\%, subject to an efficiency constraint (latency reduction per shed request) that discourages shedding without a corresponding latency benefit.
The entire loop, 10 iterations comprising over 60 simulation runs across agent-designed workloads, completed in approximately 75 minutes of wall-clock time, of which BLIS consumed 7.5 minutes, the rest in agent reasoning and policy integration.

\smallskip\noindent\textbf{The discovered policy.}
\label{subsec:discovered-policy}
Early iterations revealed a tension: aggressive shedding reduced latency at moderate load but shed excessively at low load.
The agent identified that the shape of the shedding curve, rather than signal smoothing, separates these regimes.
The resulting policy, the probabilistic admitter, is defined per request.
Requests in critical tiers are always admitted.
A sheddable request is shed with probability
\begin{equation}
  p_{\text{shed}}(s) = \min\!\left(300\, s^5,\ 1\right).
\end{equation}
The high exponent concentrates shedding near saturation: approximately 1\% at $s{=}0.12$, 29\% at $s{=}0.25$, and full shedding as $s$ approaches 0.32, replacing the binary threshold with a graded response.

In simulation, across the agent-designed workload suite (blinded validation workloads, to avoid overfitting to a specific traffic mix), the probabilistic admitter reduced critical-tier TTFT by 17--37\% (mean) and 19--44\% (P90), and E2E latency by 17--46\% (mean and P90), relative to the default policy.

\smallskip\noindent\textbf{Real-cluster validation.}
\label{subsec:real-validation}
We validated the policy on Qwen3-14B served by vLLM~\cite{kwon2023vllm} on four NVIDIA H100-80GB SXM GPUs routed through llm-d, comparing the default policy against the probabilistic admitter on two workload profiles from the llm-d benchmarking testbed~\cite{llmdbenchmark2024} (\emph{code completion}: 2K input / 128 output tokens, 30\% critical / 70\% sheddable; \emph{interactive chat}: 4K input / 1K output tokens, 80\% critical / 20\% sheddable), each at two load levels.

On the sheddable-heavy \emph{code-completion} workload, the probabilistic admitter reduced critical-tier TTFT by 74\% (mean) and 82\% (P90), and E2E latency by 50\% (mean and P90), pooled across load levels (82\% at near-saturation; negligible at low load where contention is absent).
The shed rate increased by 6.8 percentage points and throughput decreased by 14\%; the shed requests would otherwise incur long latency under overload.
On the critical-heavy \emph{interactive-chat} workload, the policy produced no meaningful change (${\approx}$2\% TTFT, 1\% E2E), confirming no regression.
The simulation and validation workloads differ by design. What matters is relative ranking: the policy that BLIS ranked above the default also improved critical-tier latency on real hardware.

\smallskip\noindent\textbf{Outcome.}
The probabilistic admitter was proposed to the llm-d open-source project and merged into the router as a configurable plugin with the simulation-derived parameters as defaults.\footnote{\url{https://github.com/llm-d/llm-d-router/tree/main/pkg/epp/framework/plugins/requestcontrol/admitter/probabilisticadmitter}}

\subsection{Flow Control: Soft-Reflective Ceiling}
\label{subsec:soft-reflective}

The probabilistic admitter addresses overload by shedding sheddable requests entirely.
A complementary problem arises when the operator wants to preserve all requests but protect critical-band latency by throttling dispatch as saturation rises.
The \emph{soft-reflective ceiling} policy implements this.

The policy assigns each priority band $i \in \{0, \ldots, N{-}1\}$ (band 0 being critical) a ceiling
\begin{equation}
  c_i(s) = 1 - \frac{i \cdot s}{N - 1},
\end{equation}
where $s$ is the cluster saturation signal defined in \S\ref{subsec:admission-problem}.
Band~0 is never gated ($c_0 = 1$ for all $s$).
When saturation exceeds a band's ceiling, dispatch to that band is rate-limited to every $\lceil s/(1{-}s) \rceil$-th scheduling tick.
The policy was again evolved using the BLIS-backed workflow described in \S\ref{subsec:policy-search}.

In simulation, across three workload profiles at near-capacity load, the soft-reflective ceiling reduced critical-class TTFT by 34--94\% (mean) and E2E latency by 32--68\% (mean) relative to the default constant-ceiling baseline, with negligible effect below the throughput knee.

\smallskip\noindent\textbf{Real-cluster validation.}
We validated the policy using the same methodology as \S\ref{subsec:real-validation}: Qwen3-14B on H100 GPUs routed through llm-d, comparing no-policy baseline against the soft-reflective ceiling.
The deployment uses 2$\times$H100 (sustainable throughput ${\approx}17$ req/s per instance, defined as the knee of the latency-load curve), with 80\% critical / 20\% sheddable traffic.
We evaluate three workload profiles spanning distinct output-length regimes: \emph{code generation} (2K input / 128 output tokens, 4--24\,rps), \emph{interactive chat} (4K input / 1K output tokens, 5--40\,rps), and \emph{reasoning} (1K input / 1.5K output tokens, 0.2--1.2\,rps).
We add a constant-ceiling control (ceiling fixed at 1.0) to isolate the algorithm's marginal contribution from router overhead.
The benefit is regime-dependent: near the throughput knee (${\approx}1\times$ sustainable load), the policy reduces E2E P50 by 32--48\% relative to no-policy baseline; below saturation it is a no-op or adds slight gating delay (+5--6\% E2E on chat, +31\% on reasoning at 0.2\,rps); under heavy overload (reasoning at 1.2\,rps), it trades TTFT (+88\%) for TPOT ($-$32\%), increasing E2E by 30\%.
The marginal contribution beyond the router-only control is 11\% TPOT and 6\% E2E across the overloaded subset.
Table~\ref{tab:soft-reflective} (Appendix~\ref{app:soft-reflective}) reports per-workload results; hardware validation confirms the simulated gains transfer to real serving.

\smallskip\noindent\textbf{Outcome.}
Like the probabilistic admitter, the soft-reflective ceiling policy was merged into the llm-d router as a configurable flow-control plugin.\footnote{\url{https://github.com/llm-d/llm-d-router/tree/main/pkg/epp/framework/plugins/flowcontrol/usagelimits/softreflectiveceiling}}

%% file: limitations.tex
\section{Discussion and Limitations}
\label{sec:limitations}

The accuracy study (\S\ref{sec:eval}) characterizes where BLIS's predictions degrade; this section addresses the simulator's broader limitations to be addressed in future work. Most limitations below are matters of coverage rather than of design: each unmodeled feature maps to a specific model term or module (\S\ref{sec:design}), so adding it extends BLIS rather than restructuring it.

\smallskip\noindent\textbf{Validation scale.}
The evidence in this paper spans two scales.
The latency-model training (\S\ref{subsec:fitting}) and the accuracy validation (\S\ref{sec:eval}) are single-instance: each experiment fits or measures one vLLM instance of up to 8 GPUs under tensor parallelism.
The configuration search (\S\ref{sec:config}) and both policy case studies (\S\ref{sec:evolution}) instead run multi-instance, on real llm-d deployments of up to 8 H100 GPUs across multiple replicas, with the probabilistic admitter validated on 4$\times$H100 and the soft-reflective ceiling on 2$\times$H100.
The case-study validations measure \emph{relative} policy improvement rather than absolute latency, which is the quantity a policy search requires (\S\ref{subsec:real-validation}), and provide preliminary evidence that policies discovered in BLIS transfer to multi-instance serving.
BLIS's absolute prediction error at multi-node scale, where cross-node interconnect and placement effects come into play, remains to be measured, and we leave this to future work.

\smallskip\noindent\textbf{Interconnect and parallelism.}
BLIS reduces communication to a bandwidth-based analytical term (\S\ref{sec:design}) that omits three effects.
Topology is invisible, so peers sharing an NVLink domain, a rack, or a cross-rack path are modeled identically.
Contention is under-modeled, since each transfer is assumed to see full bandwidth; this is why the all-reduce term under-estimates NVLink contention at higher tensor-parallel degrees (\S\ref{subsec:failure-modes}) and would diverge further on shared cross-node links.
Latency variability is absent, as a deterministic constant cannot reproduce jitter from queueing or congestion.
A topology- and contention-aware term would close these, with larger gains across nodes.
In terms of parallelism, the main gap is expert parallelism, where the all-reduce term does not capture all-to-all dispatch and would enter as a separate term (pipeline and hybrid parallelism~\cite{huang2019gpipe,narayanan2019pipedream} are likewise unmodeled).

\smallskip\noindent\textbf{Serving-technique coverage.}
BLIS models the subsystems most relevant to policy and configuration (scheduling, batching, KV-cache management, admission control), but not multi-adapter (LoRA) serving, disaggregated prefill/decode~\cite{zhong2024distserve} with cross-pool KV-cache transfer, or recent algorithmic optimizations.
Speculative decoding~\cite{leviathan2023speculative} replaces the one-token-per-step assumption with a draft-and-verify loop; KV-cache compression shrinks the per-token cache, rescaling the decode memory-bandwidth term ($T^{\text{dc}}_{\text{kv}}$, \S\ref{sec:design}); LoRA maps to an adapter-aware batch-formation module; and disaggregation maps to separate prefill and decode pools linked by a transfer term.

\smallskip\noindent\textbf{Model, hardware, and workload scope.}
Our evaluation covers dense and mixture-of-experts transformers on NVIDIA GPUs (H100, A100, L40S), drawn largely from a few popular model families (Llama, Qwen, Mistral), under ServeGen and Azure traces (\S\ref{subsec:setup}).
Within this family BLIS generalizes to unseen models and GPUs without re-fitting, since the basis functions read their inputs (FLOPs, memory traffic, layer and head counts) from the HuggingFace \texttt{config.json} and GPU datasheet (\S\ref{subsec:fitting}); a new architecture adds a basis term while existing coefficients stay valid (Appendix~\ref{app:training}).
Generalization may weaken where those inputs no longer capture execution: state-space models~\cite{gu2023mamba} and RWKV~\cite{peng2023rwkv} carry recurrent state in place of the KV cache, and multi-head latent attention~\cite{deepseek2024v2} reshapes the per-token cache, neither of which the current decode term models.
Non-NVIDIA accelerators (AMD GPUs, TPUs) are untested, and the NVIDIA-fit overhead terms would likely need re-fitting per vendor.
Workloads with extreme characteristics absent from our traces (e.g., very long multi-turn conversations or bursty order-of-magnitude spikes) may also exhibit different error profiles.

\smallskip\noindent\textbf{Platform drift.}
BLIS mirrors the scheduling semantics of a specific vLLM version (\S\ref{subsec:setup}) and the distributed inference logic of llm-d, so its modeled physics are tied to those implementations and must be updated as they evolve. Modular policy interfaces (\S\ref{sec:design}) let components update independently.

\smallskip\noindent\textbf{Saturated regimes.}
BLIS's latency model is calibrated for the region where the system is loaded but not deeply saturated; under heavy overload, small perturbations in arrival or service times produce disproportionate queueing effects that no deterministic simulator can predict precisely.
The consequence is bounded: search and policy evolution seek operating points that \emph{avoid} saturation, where BLIS reliably separates configurations that meet SLOs from those that do not.

%% file: accuracy_runtime.tex
\begin{tabular}{lrr}
\toprule
Simulator & Median Runtime (s) & Speedup vs.\ Real \\
\midrule
BLIS & 5.7 & 212$\times$ \\
Vidur & 8.6 & 141$\times$ \\
LLM-Optimizer & 0.1 & 22{,}809$\times$ \\
AIConfigurator & 2.1 & 586$\times$ \\
LLMServingSim & 604.2 & 0.1$\times$ \\
\bottomrule
\end{tabular}

%% file: main.bbl

\begin{thebibliography}{52}


\ifx \showCODEN    \undefined \def \showCODEN     #1{\unskip}     \fi
\ifx \showISBNx    \undefined \def \showISBNx     #1{\unskip}     \fi
\ifx \showISBNxiii \undefined \def \showISBNxiii  #1{\unskip}     \fi
\ifx \showISSN     \undefined \def \showISSN      #1{\unskip}     \fi
\ifx \showLCCN     \undefined \def \showLCCN      #1{\unskip}     \fi
\ifx \shownote     \undefined \def \shownote      #1{#1}          \fi
\ifx \showarticletitle \undefined \def \showarticletitle #1{#1}   \fi
\ifx \showURL      \undefined \def \showURL       {\relax}        \fi
\providecommand\bibfield[2]{#2}
\providecommand\bibinfo[2]{#2}
\providecommand\natexlab[1]{#1}
\providecommand\showeprint[2][]{arXiv:#2}

\bibitem[Agrawal et~al\mbox{.}(2024a)]%
        {agrawal2024vidur}
\bibfield{author}{\bibinfo{person}{Amey Agrawal}, \bibinfo{person}{Nitin
  Kedia}, \bibinfo{person}{Jayashree Mohan}, \bibinfo{person}{Ashish Panwar},
  \bibinfo{person}{Nipun Kwatra}, \bibinfo{person}{Bhargav~S. Gulavani},
  \bibinfo{person}{Ramachandran Ramjee}, {and} \bibinfo{person}{Alexey
  Tumanov}.} \bibinfo{year}{2024}\natexlab{a}.
\newblock \showarticletitle{VIDUR: A LARGE-SCALE SIMULATION FRAMEWORK FOR LLM
  INFERENCE}. In \bibinfo{booktitle}{\emph{Proceedings of Machine Learning and
  Systems}}, \bibfield{editor}{\bibinfo{person}{P.~Gibbons},
  \bibinfo{person}{G.~Pekhimenko}, {and} \bibinfo{person}{C.~De Sa}} (Eds.),
  Vol.~\bibinfo{volume}{6}. \bibinfo{pages}{351--366}.
\newblock
\urldef\tempurl%
\url{https://proceedings.mlsys.org/paper_files/paper/2024/file/b74a8de47d2b3c928360e0a011f48351-Paper-Conference.pdf}
\showURL{%
\tempurl}


\bibitem[Agrawal et~al\mbox{.}(2024b)]%
        {agrawal2024sarathi}
\bibfield{author}{\bibinfo{person}{Amey Agrawal}, \bibinfo{person}{Nitin
  Kedia}, \bibinfo{person}{Ashish Panwar}, \bibinfo{person}{Jayashree Mohan},
  \bibinfo{person}{Nipun Kwatra}, \bibinfo{person}{Bhargav~S. Gulavani},
  \bibinfo{person}{Alexey Tumanov}, {and} \bibinfo{person}{Ramachandran
  Ramjee}.} \bibinfo{year}{2024}\natexlab{b}.
\newblock \showarticletitle{Taming throughput-latency tradeoff in LLM inference
  with sarathi-serve}. In \bibinfo{booktitle}{\emph{Proceedings of the 18th
  USENIX Conference on Operating Systems Design and Implementation}} (Santa
  Clara, CA, USA) \emph{(\bibinfo{series}{OSDI'24})}.
  \bibinfo{publisher}{USENIX Association}, \bibinfo{address}{USA}, Article
  \bibinfo{articleno}{7}, \bibinfo{numpages}{18}~pages.
\newblock
\showISBNx{978-1-939133-40-3}


\bibitem[{AI-native Systems Research}(2025)]%
        {nous2025}
\bibfield{author}{\bibinfo{person}{{AI-native Systems Research}}.}
  \bibinfo{year}{2025}\natexlab{}.
\newblock \bibinfo{title}{Nous: A Framework That Runs the Scientific Method on
  Software Systems}.
\newblock
  \bibinfo{howpublished}{\url{https://github.com/AI-native-Systems-Research/agentic-strategy-evolution}}.
\newblock


\bibitem[Akiba et~al\mbox{.}(2019)]%
        {akiba2019optuna}
\bibfield{author}{\bibinfo{person}{Takuya Akiba}, \bibinfo{person}{Shotaro
  Sano}, \bibinfo{person}{Toshihiko Yanase}, \bibinfo{person}{Takeru Ohta},
  {and} \bibinfo{person}{Masanori Koyama}.} \bibinfo{year}{2019}\natexlab{}.
\newblock \showarticletitle{Optuna: A Next-generation Hyperparameter
  Optimization Framework}. In \bibinfo{booktitle}{\emph{Proceedings of the 25th
  ACM SIGKDD International Conference on Knowledge Discovery \& Data Mining}}
  (Anchorage, AK, USA) \emph{(\bibinfo{series}{KDD '19})}.
  \bibinfo{publisher}{Association for Computing Machinery},
  \bibinfo{address}{New York, NY, USA}, \bibinfo{pages}{2623–2631}.
\newblock
\showISBNx{9781450362016}
\href{https://doi.org/10.1145/3292500.3330701}{doi:\nolinkurl{10.1145/3292500.3330701}}


\bibitem[{Alibaba}(2025)]%
        {infersim2025}
\bibfield{author}{\bibinfo{person}{{Alibaba}}.}
  \bibinfo{year}{2025}\natexlab{}.
\newblock \bibinfo{title}{{InferSim}: A Lightweight {LLM} Inference Performance
  Simulator}.
\newblock \bibinfo{howpublished}{\url{https://github.com/alibaba/InferSim}}.
\newblock


\bibitem[Bambhaniya et~al\mbox{.}(2025)]%
        {bambhaniya2024genz}
\bibfield{author}{\bibinfo{person}{Abhimanyu Bambhaniya},
  \bibinfo{person}{Ritik Raj}, \bibinfo{person}{Geonhwa Jeong},
  \bibinfo{person}{Souvik Kundu}, \bibinfo{person}{Sudarshan Srinivasan},
  \bibinfo{person}{Suvinay Subramanian}, \bibinfo{person}{Midhilesh
  Elavazhagan}, \bibinfo{person}{Madhu Kumar}, {and} \bibinfo{person}{Tushar
  Krishna}.} \bibinfo{year}{2025}\natexlab{}.
\newblock \bibinfo{title}{Demystifying AI Platform Design for Distributed
  Inference of Next-Generation LLM models}.
\newblock
\showeprint[arxiv]{2406.01698}~[cs.AR]
\urldef\tempurl%
\url{https://arxiv.org/abs/2406.01698}
\showURL{%
\tempurl}


\bibitem[Banks et~al\mbox{.}(2010)]%
        {banks2010des}
\bibfield{author}{\bibinfo{person}{Jerry Banks}, \bibinfo{person}{\{John S.\}
  Carson}, \bibinfo{person}{\{Barry L.\} Nelson}, {and}
  \bibinfo{person}{\{David Malcolm\} Nicol}.} \bibinfo{year}{2010}\natexlab{}.
\newblock \bibinfo{booktitle}{\emph{Discrete-Event System Simulation}
  (\bibinfo{edition}{5} ed.)}.
\newblock \bibinfo{publisher}{Prentice-Hall}.
\newblock
\showISBNx{9780136062127}


\bibitem[{BentoML}(2024)]%
        {bentoml2024llmoptimizer}
\bibfield{author}{\bibinfo{person}{{BentoML}}.}
  \bibinfo{year}{2024}\natexlab{}.
\newblock \bibinfo{title}{llm-optimizer: {LLM} Inference Optimization and
  Benchmarking Tool}.
\newblock
  \bibinfo{howpublished}{\url{https://github.com/bentoml/llm-optimizer}}.
\newblock


\bibitem[Bergstra et~al\mbox{.}(2011)]%
        {bergstra2011tpe}
\bibfield{author}{\bibinfo{person}{James Bergstra}, \bibinfo{person}{R\'{e}mi
  Bardenet}, \bibinfo{person}{Yoshua Bengio}, {and} \bibinfo{person}{Bal\'{a}zs
  K\'{e}gl}.} \bibinfo{year}{2011}\natexlab{}.
\newblock \showarticletitle{Algorithms for Hyper-Parameter Optimization}. In
  \bibinfo{booktitle}{\emph{Advances in Neural Information Processing
  Systems}}, \bibfield{editor}{\bibinfo{person}{J.~Shawe-Taylor},
  \bibinfo{person}{R.~Zemel}, \bibinfo{person}{P.~Bartlett},
  \bibinfo{person}{F.~Pereira}, {and} \bibinfo{person}{K.~Weinberger}} (Eds.),
  Vol.~\bibinfo{volume}{24}. \bibinfo{publisher}{Curran Associates, Inc.}
\newblock
\urldef\tempurl%
\url{https://proceedings.neurips.cc/paper_files/paper/2011/file/86e8f7ab32cfd12577bc2619bc635690-Paper.pdf}
\showURL{%
\tempurl}


\bibitem[Cheng and Nguyen(2026)]%
        {cheng2026milp}
\bibfield{author}{\bibinfo{person}{Jiaming Cheng} {and}
  \bibinfo{person}{Duong~Tung Nguyen}.} \bibinfo{year}{2026}\natexlab{}.
\newblock \bibinfo{title}{Scalable Joint Resource Allocation for
  SLO-Constrained LLM Inference in Heterogeneous GPU Clouds}.
\newblock
\showeprint[arxiv]{2604.07472}~[cs.LG]
\urldef\tempurl%
\url{https://arxiv.org/abs/2604.07472}
\showURL{%
\tempurl}


\bibitem[chi Hsieh et~al\mbox{.}(2026)]%
        {hsieh2026flowprefill}
\bibfield{author}{\bibinfo{person}{Chia chi Hsieh}, \bibinfo{person}{Zan Zong},
  \bibinfo{person}{Xinyang Chen}, \bibinfo{person}{Jianjiang Li},
  \bibinfo{person}{Jidong Zhai}, {and} \bibinfo{person}{Lijie Wen}.}
  \bibinfo{year}{2026}\natexlab{}.
\newblock \bibinfo{title}{FlowPrefill: Decoupling Preemption from Prefill
  Scheduling Granularity to Mitigate Head-of-Line Blocking in LLM Serving}.
\newblock
\showeprint[arxiv]{2602.16603}~[cs.DC]
\urldef\tempurl%
\url{https://arxiv.org/abs/2602.16603}
\showURL{%
\tempurl}


\bibitem[Cho et~al\mbox{.}(2026a)]%
        {cho2026pascal}
\bibfield{author}{\bibinfo{person}{Eunyeong Cho}, \bibinfo{person}{Jehyeon
  Bang}, \bibinfo{person}{Ranggi Hwang}, {and} \bibinfo{person}{Minsoo Rhu}.}
  \bibinfo{year}{2026}\natexlab{a}.
\newblock \showarticletitle{{ PASCAL: A Phase-Aware Scheduling Algorithm for
  Serving Reasoning-based Large Language Models }}. In
  \bibinfo{booktitle}{\emph{2026 IEEE International Symposium on High
  Performance Computer Architecture (HPCA)}}. \bibinfo{publisher}{IEEE Computer
  Society}, \bibinfo{address}{Los Alamitos, CA, USA}, \bibinfo{pages}{1--16}.
\newblock
\href{https://doi.org/10.1109/HPCA68181.2026.11408493}{doi:\nolinkurl{10.1109/HPCA68181.2026.11408493}}


\bibitem[Cho et~al\mbox{.}(2026b)]%
        {cho2026llmservingsim2}
\bibfield{author}{\bibinfo{person}{Jaehong Cho}, \bibinfo{person}{Hyunmin
  Choi}, \bibinfo{person}{Guseul Heo}, {and} \bibinfo{person}{Jongse Park}.}
  \bibinfo{year}{2026}\natexlab{b}.
\newblock \showarticletitle{{LLMServingSim 2.0: A Unified Simulator for
  Heterogeneous and Disaggregated LLM Serving Infrastructure}}. In
  \bibinfo{booktitle}{\emph{2026 IEEE International Symposium on Performance
  Analysis of Systems and Software (ISPASS)}}. \bibinfo{pages}{1--14}.
\newblock
\href{https://doi.org/10.1109/ISPASS69572.2026.00012}{doi:\nolinkurl{10.1109/ISPASS69572.2026.00012}}


\bibitem[Cho et~al\mbox{.}(2024)]%
        {cho2024llmservingsim}
\bibfield{author}{\bibinfo{person}{Jaehong Cho}, \bibinfo{person}{Minsu Kim},
  \bibinfo{person}{Hyunmin Choi}, \bibinfo{person}{Guseul Heo}, {and}
  \bibinfo{person}{Jongse Park}.} \bibinfo{year}{2024}\natexlab{}.
\newblock \showarticletitle{{LLMServingSim: A HW/SW Co-Simulation
  Infrastructure for LLM Inference Serving at Scale}}. In
  \bibinfo{booktitle}{\emph{2024 IEEE International Symposium on Workload
  Characterization (IISWC)}}. \bibinfo{pages}{15--29}.
\newblock
\href{https://doi.org/10.1109/IISWC63097.2024.00012}{doi:\nolinkurl{10.1109/IISWC63097.2024.00012}}


\bibitem[Dao et~al\mbox{.}(2022)]%
        {dao2022flashattention}
\bibfield{author}{\bibinfo{person}{Tri Dao}, \bibinfo{person}{Daniel~Y. Fu},
  \bibinfo{person}{Stefano Ermon}, \bibinfo{person}{Atri Rudra}, {and}
  \bibinfo{person}{Christopher R\'{e}}.} \bibinfo{year}{2022}\natexlab{}.
\newblock \showarticletitle{FLASHATTENTION: fast and memory-efficient exact
  attention with IO-awareness}. In \bibinfo{booktitle}{\emph{Proceedings of the
  36th International Conference on Neural Information Processing Systems}} (New
  Orleans, LA, USA) \emph{(\bibinfo{series}{NIPS '22})}.
  \bibinfo{publisher}{Curran Associates Inc.}, \bibinfo{address}{Red Hook, NY,
  USA}, Article \bibinfo{articleno}{1189}, \bibinfo{numpages}{16}~pages.
\newblock
\showISBNx{9781713871088}


\bibitem[DeepSeek-AI et~al\mbox{.}(2024)]%
        {deepseek2024v2}
\bibfield{author}{\bibinfo{person}{DeepSeek-AI}, \bibinfo{person}{Aixin Liu},
  \bibinfo{person}{Bei Feng}, \bibinfo{person}{Bin Wang},
  \bibinfo{person}{Bingxuan Wang}, \bibinfo{person}{Bo Liu},
  \bibinfo{person}{Chenggang Zhao}, \bibinfo{person}{Chengqi Dengr},
  \bibinfo{person}{Chong Ruan}, \bibinfo{person}{Damai Dai},
  \bibinfo{person}{Daya Guo}, \bibinfo{person}{Dejian Yang},
  \bibinfo{person}{Deli Chen}, \bibinfo{person}{Dongjie Ji},
  \bibinfo{person}{Erhang Li}, \bibinfo{person}{Fangyun Lin},
  \bibinfo{person}{Fuli Luo}, \bibinfo{person}{Guangbo Hao},
  \bibinfo{person}{Guanting Chen}, \bibinfo{person}{Guowei Li},
  \bibinfo{person}{H. Zhang}, \bibinfo{person}{Hanwei Xu}, \bibinfo{person}{Hao
  Yang}, \bibinfo{person}{Haowei Zhang}, \bibinfo{person}{Honghui Ding},
  \bibinfo{person}{Huajian Xin}, \bibinfo{person}{Huazuo Gao},
  \bibinfo{person}{Hui Li}, \bibinfo{person}{Hui Qu}, \bibinfo{person}{J.~L.
  Cai}, \bibinfo{person}{Jian Liang}, \bibinfo{person}{Jianzhong Guo},
  \bibinfo{person}{Jiaqi Ni}, \bibinfo{person}{Jiashi Li}, \bibinfo{person}{Jin
  Chen}, \bibinfo{person}{Jingyang Yuan}, \bibinfo{person}{Junjie Qiu},
  \bibinfo{person}{Junxiao Song}, \bibinfo{person}{Kai Dong},
  \bibinfo{person}{Kaige Gao}, \bibinfo{person}{Kang Guan},
  \bibinfo{person}{Lean Wang}, \bibinfo{person}{Lecong Zhang},
  \bibinfo{person}{Lei Xu}, \bibinfo{person}{Leyi Xia}, \bibinfo{person}{Liang
  Zhao}, \bibinfo{person}{Liyue Zhang}, \bibinfo{person}{Meng Li},
  \bibinfo{person}{Miaojun Wang}, \bibinfo{person}{Mingchuan Zhang},
  \bibinfo{person}{Minghua Zhang}, \bibinfo{person}{Minghui Tang},
  \bibinfo{person}{Mingming Li}, \bibinfo{person}{Ning Tian},
  \bibinfo{person}{Panpan Huang}, \bibinfo{person}{Peiyi Wang},
  \bibinfo{person}{Peng Zhang}, \bibinfo{person}{Qihao Zhu},
  \bibinfo{person}{Qinyu Chen}, \bibinfo{person}{Qiushi Du},
  \bibinfo{person}{R.~J. Chen}, \bibinfo{person}{R.~L. Jin},
  \bibinfo{person}{Ruiqi Ge}, \bibinfo{person}{Ruizhe Pan},
  \bibinfo{person}{Runxin Xu}, \bibinfo{person}{Ruyi Chen},
  \bibinfo{person}{S.~S. Li}, \bibinfo{person}{Shanghao Lu},
  \bibinfo{person}{Shangyan Zhou}, \bibinfo{person}{Shanhuang Chen},
  \bibinfo{person}{Shaoqing Wu}, \bibinfo{person}{Shengfeng Ye},
  \bibinfo{person}{Shirong Ma}, \bibinfo{person}{Shiyu Wang},
  \bibinfo{person}{Shuang Zhou}, \bibinfo{person}{Shuiping Yu},
  \bibinfo{person}{Shunfeng Zhou}, \bibinfo{person}{Size Zheng},
  \bibinfo{person}{T. Wang}, \bibinfo{person}{Tian Pei}, \bibinfo{person}{Tian
  Yuan}, \bibinfo{person}{Tianyu Sun}, \bibinfo{person}{W.~L. Xiao},
  \bibinfo{person}{Wangding Zeng}, \bibinfo{person}{Wei An},
  \bibinfo{person}{Wen Liu}, \bibinfo{person}{Wenfeng Liang},
  \bibinfo{person}{Wenjun Gao}, \bibinfo{person}{Wentao Zhang},
  \bibinfo{person}{X.~Q. Li}, \bibinfo{person}{Xiangyue Jin},
  \bibinfo{person}{Xianzu Wang}, \bibinfo{person}{Xiao Bi},
  \bibinfo{person}{Xiaodong Liu}, \bibinfo{person}{Xiaohan Wang},
  \bibinfo{person}{Xiaojin Shen}, \bibinfo{person}{Xiaokang Chen},
  \bibinfo{person}{Xiaosha Chen}, \bibinfo{person}{Xiaotao Nie},
  \bibinfo{person}{Xiaowen Sun}, \bibinfo{person}{Xiaoxiang Wang},
  \bibinfo{person}{Xin Liu}, \bibinfo{person}{Xin Xie},
  \bibinfo{person}{Xingkai Yu}, \bibinfo{person}{Xinnan Song},
  \bibinfo{person}{Xinyi Zhou}, \bibinfo{person}{Xinyu Yang},
  \bibinfo{person}{Xuan Lu}, \bibinfo{person}{Xuecheng Su}, \bibinfo{person}{Y.
  Wu}, \bibinfo{person}{Y.~K. Li}, \bibinfo{person}{Y.~X. Wei},
  \bibinfo{person}{Y.~X. Zhu}, \bibinfo{person}{Yanhong Xu},
  \bibinfo{person}{Yanping Huang}, \bibinfo{person}{Yao Li},
  \bibinfo{person}{Yao Zhao}, \bibinfo{person}{Yaofeng Sun},
  \bibinfo{person}{Yaohui Li}, \bibinfo{person}{Yaohui Wang},
  \bibinfo{person}{Yi Zheng}, \bibinfo{person}{Yichao Zhang},
  \bibinfo{person}{Yiliang Xiong}, \bibinfo{person}{Yilong Zhao},
  \bibinfo{person}{Ying He}, \bibinfo{person}{Ying Tang},
  \bibinfo{person}{Yishi Piao}, \bibinfo{person}{Yixin Dong},
  \bibinfo{person}{Yixuan Tan}, \bibinfo{person}{Yiyuan Liu},
  \bibinfo{person}{Yongji Wang}, \bibinfo{person}{Yongqiang Guo},
  \bibinfo{person}{Yuchen Zhu}, \bibinfo{person}{Yuduan Wang},
  \bibinfo{person}{Yuheng Zou}, \bibinfo{person}{Yukun Zha},
  \bibinfo{person}{Yunxian Ma}, \bibinfo{person}{Yuting Yan},
  \bibinfo{person}{Yuxiang You}, \bibinfo{person}{Yuxuan Liu},
  \bibinfo{person}{Z.~Z. Ren}, \bibinfo{person}{Zehui Ren},
  \bibinfo{person}{Zhangli Sha}, \bibinfo{person}{Zhe Fu},
  \bibinfo{person}{Zhen Huang}, \bibinfo{person}{Zhen Zhang},
  \bibinfo{person}{Zhenda Xie}, \bibinfo{person}{Zhewen Hao},
  \bibinfo{person}{Zhihong Shao}, \bibinfo{person}{Zhiniu Wen},
  \bibinfo{person}{Zhipeng Xu}, \bibinfo{person}{Zhongyu Zhang},
  \bibinfo{person}{Zhuoshu Li}, \bibinfo{person}{Zihan Wang},
  \bibinfo{person}{Zihui Gu}, \bibinfo{person}{Zilin Li}, {and}
  \bibinfo{person}{Ziwei Xie}.} \bibinfo{year}{2024}\natexlab{}.
\newblock \bibinfo{title}{DeepSeek-V2: A Strong, Economical, and Efficient
  Mixture-of-Experts Language Model}.
\newblock
\showeprint[arxiv]{2405.04434}~[cs.CL]
\urldef\tempurl%
\url{https://arxiv.org/abs/2405.04434}
\showURL{%
\tempurl}


\bibitem[Gu and Dao(2024)]%
        {gu2023mamba}
\bibfield{author}{\bibinfo{person}{Albert Gu} {and} \bibinfo{person}{Tri Dao}.}
  \bibinfo{year}{2024}\natexlab{}.
\newblock \bibinfo{title}{Mamba: Linear-Time Sequence Modeling with Selective
  State Spaces}.
\newblock
\showeprint[arxiv]{2312.00752}~[cs.LG]
\urldef\tempurl%
\url{https://arxiv.org/abs/2312.00752}
\showURL{%
\tempurl}


\bibitem[Hansen et~al\mbox{.}(2003)]%
        {hansen2003cmaes}
\bibfield{author}{\bibinfo{person}{Nikolaus Hansen},
  \bibinfo{person}{Sibylle~D. Müller}, {and} \bibinfo{person}{Petros
  Koumoutsakos}.} \bibinfo{year}{2003}\natexlab{}.
\newblock \showarticletitle{Reducing the Time Complexity of the Derandomized
  Evolution Strategy with Covariance Matrix Adaptation (CMA-ES)}.
\newblock \bibinfo{journal}{\emph{Evolutionary Computation}}
  \bibinfo{volume}{11}, \bibinfo{number}{1} (\bibinfo{year}{2003}),
  \bibinfo{pages}{1--18}.
\newblock
\href{https://doi.org/10.1162/106365603321828970}{doi:\nolinkurl{10.1162/106365603321828970}}


\bibitem[Huang et~al\mbox{.}(2019)]%
        {huang2019gpipe}
\bibfield{author}{\bibinfo{person}{Yanping Huang}, \bibinfo{person}{Youlong
  Cheng}, \bibinfo{person}{Ankur Bapna}, \bibinfo{person}{Orhan Firat},
  \bibinfo{person}{Mia~Xu Chen}, \bibinfo{person}{Dehao Chen},
  \bibinfo{person}{HyoukJoong Lee}, \bibinfo{person}{Jiquan Ngiam},
  \bibinfo{person}{Quoc~V. Le}, \bibinfo{person}{Yonghui Wu}, {and}
  \bibinfo{person}{Zhifeng Chen}.} \bibinfo{year}{2019}\natexlab{}.
\newblock \bibinfo{booktitle}{\emph{GPipe: efficient training of giant neural
  networks using pipeline parallelism}}.
\newblock \bibinfo{publisher}{Curran Associates Inc.}, \bibinfo{address}{Red
  Hook, NY, USA}.
\newblock


\bibitem[Kiefer(1953)]%
        {kiefer1953golden}
\bibfield{author}{\bibinfo{person}{J. Kiefer}.}
  \bibinfo{year}{1953}\natexlab{}.
\newblock \showarticletitle{Sequential Minimax Search for a Maximum}.
\newblock \bibinfo{journal}{\emph{Proc. Amer. Math. Soc.}} \bibinfo{volume}{4},
  \bibinfo{number}{3} (\bibinfo{year}{1953}), \bibinfo{pages}{502--506}.
\newblock
\showISSN{00029939, 10886826}
\urldef\tempurl%
\url{http://www.jstor.org/stable/2032161}
\showURL{%
\tempurl}


\bibitem[Korthikanti et~al\mbox{.}(2022)]%
        {korthikanti2023megatron}
\bibfield{author}{\bibinfo{person}{Vijay Korthikanti}, \bibinfo{person}{Jared
  Casper}, \bibinfo{person}{Sangkug Lym}, \bibinfo{person}{Lawrence McAfee},
  \bibinfo{person}{Michael Andersch}, \bibinfo{person}{Mohammad Shoeybi}, {and}
  \bibinfo{person}{Bryan Catanzaro}.} \bibinfo{year}{2022}\natexlab{}.
\newblock \bibinfo{title}{Reducing Activation Recomputation in Large
  Transformer Models}.
\newblock
\showeprint[arxiv]{2205.05198}~[cs.LG]
\urldef\tempurl%
\url{https://arxiv.org/abs/2205.05198}
\showURL{%
\tempurl}


\bibitem[{Kubernetes SIGs}(2024)]%
        {inferenceperf2024}
\bibfield{author}{\bibinfo{person}{{Kubernetes SIGs}}.}
  \bibinfo{year}{2024}\natexlab{}.
\newblock \bibinfo{title}{inference-perf: Production-Scale GenAI Inference
  Performance Benchmarking Tool}.
\newblock
  \bibinfo{howpublished}{\url{https://github.com/kubernetes-sigs/inference-perf}}.
\newblock


\bibitem[Kwon et~al\mbox{.}(2023)]%
        {kwon2023vllm}
\bibfield{author}{\bibinfo{person}{Woosuk Kwon}, \bibinfo{person}{Zhuohan Li},
  \bibinfo{person}{Siyuan Zhuang}, \bibinfo{person}{Ying Sheng},
  \bibinfo{person}{Lianmin Zheng}, \bibinfo{person}{Cody~Hao Yu},
  \bibinfo{person}{Joseph Gonzalez}, \bibinfo{person}{Hao Zhang}, {and}
  \bibinfo{person}{Ion Stoica}.} \bibinfo{year}{2023}\natexlab{}.
\newblock \showarticletitle{Efficient Memory Management for Large Language
  Model Serving with PagedAttention}. In \bibinfo{booktitle}{\emph{Proceedings
  of the 29th Symposium on Operating Systems Principles}} (Koblenz, Germany)
  \emph{(\bibinfo{series}{SOSP '23})}. \bibinfo{publisher}{Association for
  Computing Machinery}, \bibinfo{address}{New York, NY, USA},
  \bibinfo{pages}{611–626}.
\newblock
\showISBNx{9798400702297}
\href{https://doi.org/10.1145/3600006.3613165}{doi:\nolinkurl{10.1145/3600006.3613165}}


\bibitem[Lazuka et~al\mbox{.}(2024)]%
        {lazuka2024llmpilot}
\bibfield{author}{\bibinfo{person}{Malgorzata Lazuka}, \bibinfo{person}{Andreea
  Anghel}, {and} \bibinfo{person}{Thomas Parnell}.}
  \bibinfo{year}{2024}\natexlab{}.
\newblock \showarticletitle{LLM-Pilot: Characterize and Optimize Performance of
  your LLM Inference Services}. In \bibinfo{booktitle}{\emph{Proceedings of the
  International Conference for High Performance Computing, Networking, Storage,
  and Analysis}} (Atlanta, GA, USA) \emph{(\bibinfo{series}{SC '24})}.
  \bibinfo{publisher}{IEEE Press}, Article \bibinfo{articleno}{16},
  \bibinfo{numpages}{18}~pages.
\newblock
\showISBNx{9798350352917}
\href{https://doi.org/10.1109/SC41406.2024.00022}{doi:\nolinkurl{10.1109/SC41406.2024.00022}}


\bibitem[Lee et~al\mbox{.}(2025)]%
        {lee2025neusight}
\bibfield{author}{\bibinfo{person}{Seonho Lee}, \bibinfo{person}{Amar
  Phanishayee}, {and} \bibinfo{person}{Divya Mahajan}.}
  \bibinfo{year}{2025}\natexlab{}.
\newblock \showarticletitle{Forecasting GPU Performance for Deep Learning
  Training and Inference}. In \bibinfo{booktitle}{\emph{Proceedings of the 30th
  ACM International Conference on Architectural Support for Programming
  Languages and Operating Systems, Volume 1}} (Rotterdam, Netherlands)
  \emph{(\bibinfo{series}{ASPLOS '25})}. \bibinfo{publisher}{Association for
  Computing Machinery}, \bibinfo{address}{New York, NY, USA},
  \bibinfo{pages}{493–508}.
\newblock
\showISBNx{9798400706981}
\href{https://doi.org/10.1145/3669940.3707265}{doi:\nolinkurl{10.1145/3669940.3707265}}


\bibitem[Leviathan et~al\mbox{.}(2023)]%
        {leviathan2023speculative}
\bibfield{author}{\bibinfo{person}{Yaniv Leviathan}, \bibinfo{person}{Matan
  Kalman}, {and} \bibinfo{person}{Yossi Matias}.}
  \bibinfo{year}{2023}\natexlab{}.
\newblock \showarticletitle{Fast inference from transformers via speculative
  decoding}. In \bibinfo{booktitle}{\emph{Proceedings of the 40th International
  Conference on Machine Learning}} (Honolulu, Hawaii, USA)
  \emph{(\bibinfo{series}{ICML'23})}. \bibinfo{publisher}{JMLR.org}, Article
  \bibinfo{articleno}{795}, \bibinfo{numpages}{13}~pages.
\newblock


\bibitem[Lim et~al\mbox{.}(2026)]%
        {lim2026lodestar}
\bibfield{author}{\bibinfo{person}{Gangmuk Lim}, \bibinfo{person}{Wanyu Zhao},
  \bibinfo{person}{Brighten Godfrey}, \bibinfo{person}{Jiaxin Shan},
  \bibinfo{person}{Le Xu}, {and} \bibinfo{person}{Liguang Xie}.}
  \bibinfo{year}{2026}\natexlab{}.
\newblock \bibinfo{title}{Lodestar: An Online-Learning LLM Inference Router}.
\newblock
\showeprint[arxiv]{2606.00946}~[cs.DC]
\urldef\tempurl%
\url{https://arxiv.org/abs/2606.00946}
\showURL{%
\tempurl}


\bibitem[Lin et~al\mbox{.}(2026)]%
        {lin2026acalsim}
\bibfield{author}{\bibinfo{person}{Wei-Fen Lin}, \bibinfo{person}{Jen-Chien
  Chang}, \bibinfo{person}{Yen-Po Chen}, \bibinfo{person}{Zi-Yi Tai},
  \bibinfo{person}{Yu-Chen Chang}, \bibinfo{person}{Chia-Pao Chiang},
  \bibinfo{person}{Yu-Yang Lee}, {and} \bibinfo{person}{Yu-Jie Wang}.}
  \bibinfo{year}{2026}\natexlab{}.
\newblock \showarticletitle{ACALSim: A Scalable Parallel Simulation Framework
  for High-Performance System Design Space Exploration}.
\newblock \bibinfo{journal}{\emph{Proc. ACM Meas. Anal. Comput. Syst.}}
  \bibinfo{volume}{10}, \bibinfo{number}{2}, Article \bibinfo{articleno}{28}
  (\bibinfo{date}{May} \bibinfo{year}{2026}), \bibinfo{numpages}{24}~pages.
\newblock
\href{https://doi.org/10.1145/3805626}{doi:\nolinkurl{10.1145/3805626}}


\bibitem[Lin et~al\mbox{.}(2025)]%
        {lin2024apex}
\bibfield{author}{\bibinfo{person}{Yi-Chien Lin}, \bibinfo{person}{Woosuk
  Kwon}, \bibinfo{person}{Ronald Pineda}, {and} \bibinfo{person}{Fanny~Nina
  Paravecino}.} \bibinfo{year}{2025}\natexlab{}.
\newblock \bibinfo{title}{APEX: An Extensible and Dynamism-Aware Simulator for
  Automated Parallel Execution in LLM Serving}.
\newblock
\showeprint[arxiv]{2411.17651}~[cs.DC]
\urldef\tempurl%
\url{https://arxiv.org/abs/2411.17651}
\showURL{%
\tempurl}


\bibitem[{llm-d Community}(2024a)]%
        {llmdbenchmark2024}
\bibfield{author}{\bibinfo{person}{{llm-d Community}}.}
  \bibinfo{year}{2024}\natexlab{a}.
\newblock \bibinfo{title}{{llm-d-benchmark}: End-to-End Benchmarking Harness
  for {llm-d}}.
\newblock
  \bibinfo{howpublished}{\url{https://github.com/llm-d/llm-d-benchmark}}.
\newblock


\bibitem[{llm-d Community}(2024b)]%
        {llmd2024}
\bibfield{author}{\bibinfo{person}{{llm-d Community}}.}
  \bibinfo{year}{2024}\natexlab{b}.
\newblock \bibinfo{title}{{llm-d}: Kubernetes-Native Distributed Inference
  Serving}.
\newblock \bibinfo{howpublished}{\url{https://github.com/llm-d/llm-d}}.
\newblock


\bibitem[Mei et~al\mbox{.}(2025)]%
        {mei2025helix}
\bibfield{author}{\bibinfo{person}{Yixuan Mei}, \bibinfo{person}{Yonghao
  Zhuang}, \bibinfo{person}{Xupeng Miao}, \bibinfo{person}{Juncheng Yang},
  \bibinfo{person}{Zhihao Jia}, {and} \bibinfo{person}{Rashmi Vinayak}.}
  \bibinfo{year}{2025}\natexlab{}.
\newblock \showarticletitle{Helix: Serving Large Language Models over
  Heterogeneous GPUs and Network via Max-Flow}. In
  \bibinfo{booktitle}{\emph{Proceedings of the 30th ACM International
  Conference on Architectural Support for Programming Languages and Operating
  Systems, Volume 1}} (Rotterdam, Netherlands) \emph{(\bibinfo{series}{ASPLOS
  '25})}. \bibinfo{publisher}{Association for Computing Machinery},
  \bibinfo{address}{New York, NY, USA}, \bibinfo{pages}{586–602}.
\newblock
\showISBNx{9798400706981}
\href{https://doi.org/10.1145/3669940.3707215}{doi:\nolinkurl{10.1145/3669940.3707215}}


\bibitem[{Microsoft Azure}(2024)]%
        {azure2024publicdataset}
\bibfield{author}{\bibinfo{person}{{Microsoft Azure}}.}
  \bibinfo{year}{2024}\natexlab{}.
\newblock \bibinfo{title}{{Azure Public Dataset}: {LLM} Inference Trace}.
\newblock
  \bibinfo{howpublished}{\url{https://github.com/Azure/AzurePublicDataset}}.
\newblock


\bibitem[Nam et~al\mbox{.}(2026)]%
        {nam2026lpgsim}
\bibfield{author}{\bibinfo{person}{Hyunwoo Nam}, \bibinfo{person}{Jay~Hwan
  Lee}, \bibinfo{person}{Yeonsoo Kim}, \bibinfo{person}{Mengzhao Zhang},
  \bibinfo{person}{Jeonggeun Kim}, {and} \bibinfo{person}{Bernd Burgstaller}.}
  \bibinfo{year}{2026}\natexlab{}.
\newblock \showarticletitle{LPGSim: A Lightweight Parallel GPU Simulator
  Maximizing Speed with Trustworthy Simulation}.
\newblock \bibinfo{journal}{\emph{Proc. ACM Meas. Anal. Comput. Syst.}}
  \bibinfo{volume}{10}, \bibinfo{number}{2}, Article \bibinfo{articleno}{44}
  (\bibinfo{date}{May} \bibinfo{year}{2026}), \bibinfo{numpages}{27}~pages.
\newblock
\href{https://doi.org/10.1145/3805642}{doi:\nolinkurl{10.1145/3805642}}


\bibitem[Narayanan et~al\mbox{.}(2019)]%
        {narayanan2019pipedream}
\bibfield{author}{\bibinfo{person}{Deepak Narayanan}, \bibinfo{person}{Aaron
  Harlap}, \bibinfo{person}{Amar Phanishayee}, \bibinfo{person}{Vivek
  Seshadri}, \bibinfo{person}{Nikhil~R. Devanur}, \bibinfo{person}{Gregory~R.
  Ganger}, \bibinfo{person}{Phillip~B. Gibbons}, {and} \bibinfo{person}{Matei
  Zaharia}.} \bibinfo{year}{2019}\natexlab{}.
\newblock \showarticletitle{PipeDream: generalized pipeline parallelism for DNN
  training}. In \bibinfo{booktitle}{\emph{Proceedings of the 27th ACM Symposium
  on Operating Systems Principles}} (Huntsville, Ontario, Canada)
  \emph{(\bibinfo{series}{SOSP '19})}. \bibinfo{publisher}{Association for
  Computing Machinery}, \bibinfo{address}{New York, NY, USA},
  \bibinfo{pages}{1–15}.
\newblock
\showISBNx{9781450368735}
\href{https://doi.org/10.1145/3341301.3359646}{doi:\nolinkurl{10.1145/3341301.3359646}}


\bibitem[{Neural Magic}(2024)]%
        {guidellm2024}
\bibfield{author}{\bibinfo{person}{{Neural Magic}}.}
  \bibinfo{year}{2024}\natexlab{}.
\newblock \bibinfo{title}{{GuideLLM}: Scalable Inference Benchmarking for Large
  Language Models}.
\newblock
  \bibinfo{howpublished}{\url{https://github.com/vllm-project/guidellm}}.
\newblock


\bibitem[{NVIDIA Corporation}(2024)]%
        {nvidia2024h100datasheet}
\bibfield{author}{\bibinfo{person}{{NVIDIA Corporation}}.}
  \bibinfo{year}{2024}\natexlab{}.
\newblock \bibinfo{title}{{NVIDIA H100 Tensor Core GPU} Datasheet}.
\newblock
  \bibinfo{howpublished}{\url{https://www.nvidia.com/en-us/data-center/h100/}}.
\newblock


\bibitem[Ouyang et~al\mbox{.}(2026)]%
        {2026unicache}
\bibfield{author}{\bibinfo{person}{Bei Ouyang}, \bibinfo{person}{Yifan Qiao},
  {and} \bibinfo{person}{Jiarong Xing}.} \bibinfo{year}{2026}\natexlab{}.
\newblock \showarticletitle{UniCache: Unifying Prefix Cache Eviction for
  Heterogeneous LLM Serving Workloads}.
\newblock \bibinfo{journal}{\emph{Proc. ACM Meas. Anal. Comput. Syst.}}
  \bibinfo{volume}{10}, \bibinfo{number}{2}, Article \bibinfo{articleno}{54}
  (\bibinfo{date}{May} \bibinfo{year}{2026}), \bibinfo{numpages}{27}~pages.
\newblock
\href{https://doi.org/10.1145/3805652}{doi:\nolinkurl{10.1145/3805652}}


\bibitem[Patel et~al\mbox{.}(2025)]%
        {patel2024splitwise}
\bibfield{author}{\bibinfo{person}{Pratyush Patel}, \bibinfo{person}{Esha
  Choukse}, \bibinfo{person}{Chaojie Zhang}, \bibinfo{person}{Aashaka Shah},
  \bibinfo{person}{\'{I}\~{n}igo Goiri}, \bibinfo{person}{Saeed Maleki}, {and}
  \bibinfo{person}{Ricardo Bianchini}.} \bibinfo{year}{2025}\natexlab{}.
\newblock \showarticletitle{Splitwise: Efficient Generative LLM Inference Using
  Phase Splitting}. In \bibinfo{booktitle}{\emph{Proceedings of the 51st Annual
  International Symposium on Computer Architecture}} (Buenos Aires, Argentina)
  \emph{(\bibinfo{series}{ISCA '24})}. \bibinfo{publisher}{IEEE Press},
  \bibinfo{pages}{118–132}.
\newblock
\showISBNx{9798350326581}
\href{https://doi.org/10.1109/ISCA59077.2024.00019}{doi:\nolinkurl{10.1109/ISCA59077.2024.00019}}


\bibitem[Peng et~al\mbox{.}(2023)]%
        {peng2023rwkv}
\bibfield{author}{\bibinfo{person}{Bo Peng}, \bibinfo{person}{Eric Alcaide},
  \bibinfo{person}{Quentin Anthony}, \bibinfo{person}{Alon Albalak},
  \bibinfo{person}{Samuel Arcadinho}, \bibinfo{person}{Stella Biderman},
  \bibinfo{person}{Huanqi Cao}, \bibinfo{person}{Xin Cheng},
  \bibinfo{person}{Michael Chung}, \bibinfo{person}{Matteo Grella},
  \bibinfo{person}{Kranthi~Kiran GV}, \bibinfo{person}{Xuzheng He},
  \bibinfo{person}{Haowen Hou}, \bibinfo{person}{Jiaju Lin},
  \bibinfo{person}{Przemyslaw Kazienko}, \bibinfo{person}{Jan Kocon},
  \bibinfo{person}{Jiaming Kong}, \bibinfo{person}{Bartlomiej Koptyra},
  \bibinfo{person}{Hayden Lau}, \bibinfo{person}{Krishna Sri~Ipsit Mantri},
  \bibinfo{person}{Ferdinand Mom}, \bibinfo{person}{Atsushi Saito},
  \bibinfo{person}{Guangyu Song}, \bibinfo{person}{Xiangru Tang},
  \bibinfo{person}{Bolun Wang}, \bibinfo{person}{Johan~S. Wind},
  \bibinfo{person}{Stanislaw Wozniak}, \bibinfo{person}{Ruichong Zhang},
  \bibinfo{person}{Zhenyuan Zhang}, \bibinfo{person}{Qihang Zhao},
  \bibinfo{person}{Peng Zhou}, \bibinfo{person}{Qinghua Zhou},
  \bibinfo{person}{Jian Zhu}, {and} \bibinfo{person}{Rui-Jie Zhu}.}
  \bibinfo{year}{2023}\natexlab{}.
\newblock \bibinfo{title}{RWKV: Reinventing RNNs for the Transformer Era}.
\newblock
\showeprint[arxiv]{2305.13048}~[cs.CL]
\urldef\tempurl%
\url{https://arxiv.org/abs/2305.13048}
\showURL{%
\tempurl}


\bibitem[Qiao et~al\mbox{.}(2025)]%
        {qiao2024conserve}
\bibfield{author}{\bibinfo{person}{Yifan Qiao}, \bibinfo{person}{Shu Anzai},
  \bibinfo{person}{Shan Yu}, \bibinfo{person}{Haoran Ma}, \bibinfo{person}{Shuo
  Yang}, \bibinfo{person}{Yang Wang}, \bibinfo{person}{Miryung Kim},
  \bibinfo{person}{Yongji Wu}, \bibinfo{person}{Yang Zhou},
  \bibinfo{person}{Jiarong Xing}, \bibinfo{person}{Joseph~E. Gonzalez},
  \bibinfo{person}{Ion Stoica}, {and} \bibinfo{person}{Harry Xu}.}
  \bibinfo{year}{2025}\natexlab{}.
\newblock \bibinfo{title}{ConServe: Fine-Grained GPU Harvesting for LLM Online
  and Offline Co-Serving}.
\newblock
\showeprint[arxiv]{2410.01228}~[cs.DC]
\urldef\tempurl%
\url{https://arxiv.org/abs/2410.01228}
\showURL{%
\tempurl}


\bibitem[Reddi et~al\mbox{.}(2020)]%
        {reddi2020mlperfinference}
\bibfield{author}{\bibinfo{person}{Vijay~Janapa Reddi},
  \bibinfo{person}{Christine Cheng}, \bibinfo{person}{David Kanter},
  \bibinfo{person}{Peter Mattson}, \bibinfo{person}{Guenther Schmuelling},
  \bibinfo{person}{Carole-Jean Wu}, \bibinfo{person}{Brian Anderson},
  \bibinfo{person}{Maximilien Breughe}, \bibinfo{person}{Mark Charlebois},
  \bibinfo{person}{William Chou}, \bibinfo{person}{Ramesh Chukka},
  \bibinfo{person}{Cody Coleman}, \bibinfo{person}{Sam Davis},
  \bibinfo{person}{Pan Deng}, \bibinfo{person}{Greg Diamos},
  \bibinfo{person}{Jared Duke}, \bibinfo{person}{Dave Fick},
  \bibinfo{person}{J.~Scott Gardner}, \bibinfo{person}{Itay Hubara},
  \bibinfo{person}{Sachin Idgunji}, \bibinfo{person}{Thomas~B. Jablin},
  \bibinfo{person}{Jeff Jiao}, \bibinfo{person}{Tom~St. John},
  \bibinfo{person}{Pankaj Kanwar}, \bibinfo{person}{David Lee},
  \bibinfo{person}{Jeffery Liao}, \bibinfo{person}{Anton Lokhmotov},
  \bibinfo{person}{Francisco Massa}, \bibinfo{person}{Peng Meng},
  \bibinfo{person}{Paulius Micikevicius}, \bibinfo{person}{Colin Osborne},
  \bibinfo{person}{Gennady Pekhimenko}, \bibinfo{person}{Arun Tejusve~Raghunath
  Rajan}, \bibinfo{person}{Dilip Sequeira}, \bibinfo{person}{Ashish Sirasao},
  \bibinfo{person}{Fei Sun}, \bibinfo{person}{Hanlin Tang},
  \bibinfo{person}{Michael Thomson}, \bibinfo{person}{Frank Wei},
  \bibinfo{person}{Ephrem Wu}, \bibinfo{person}{Lingjie Xu},
  \bibinfo{person}{Koichi Yamada}, \bibinfo{person}{Bing Yu},
  \bibinfo{person}{George Yuan}, \bibinfo{person}{Aaron Zhong},
  \bibinfo{person}{Peizhao Zhang}, {and} \bibinfo{person}{Yuchen Zhou}.}
  \bibinfo{year}{2020}\natexlab{}.
\newblock \bibinfo{title}{MLPerf Inference Benchmark}.
\newblock
\showeprint[arxiv]{1911.02549}~[cs.LG]
\urldef\tempurl%
\url{https://arxiv.org/abs/1911.02549}
\showURL{%
\tempurl}


\bibitem[Shah et~al\mbox{.}(2024)]%
        {shah2024flashattention3}
\bibfield{author}{\bibinfo{person}{Jay Shah}, \bibinfo{person}{Ganesh
  Bikshandi}, \bibinfo{person}{Ying Zhang}, \bibinfo{person}{Vijay Thakkar},
  \bibinfo{person}{Pradeep Ramani}, {and} \bibinfo{person}{Tri Dao}.}
  \bibinfo{year}{2024}\natexlab{}.
\newblock \showarticletitle{FlashAttention-3: fast and accurate attention with
  asynchrony and low-precision}. In \bibinfo{booktitle}{\emph{Proceedings of
  the 38th International Conference on Neural Information Processing Systems}}
  (Vancouver, BC, Canada) \emph{(\bibinfo{series}{NIPS '24})}.
  \bibinfo{publisher}{Curran Associates Inc.}, \bibinfo{address}{Red Hook, NY,
  USA}, Article \bibinfo{articleno}{2193}, \bibinfo{numpages}{28}~pages.
\newblock
\showISBNx{9798331314385}


\bibitem[Sun et~al\mbox{.}(2026)]%
        {sun2026mfs}
\bibfield{author}{\bibinfo{person}{Yijun Sun}, \bibinfo{person}{Xudong Liao},
  \bibinfo{person}{Songrun Xie}, \bibinfo{person}{Hao Chen},
  \bibinfo{person}{Han Tian}, \bibinfo{person}{Wenxue Li},
  \bibinfo{person}{Yiming Zhang}, {and} \bibinfo{person}{Kai Chen}.}
  \bibinfo{year}{2026}\natexlab{}.
\newblock \bibinfo{title}{Stage-Aware Communication Scheduling for
  Disaggregated LLM Serving}.
\newblock
\showeprint[arxiv]{2603.17456}~[cs.NI]
\urldef\tempurl%
\url{https://arxiv.org/abs/2603.17456}
\showURL{%
\tempurl}


\bibitem[Williams et~al\mbox{.}(2009)]%
        {williams2009roofline}
\bibfield{author}{\bibinfo{person}{Samuel Williams}, \bibinfo{person}{Andrew
  Waterman}, {and} \bibinfo{person}{David Patterson}.}
  \bibinfo{year}{2009}\natexlab{}.
\newblock \showarticletitle{Roofline: an insightful visual performance model
  for multicore architectures}.
\newblock \bibinfo{journal}{\emph{Commun. ACM}} \bibinfo{volume}{52},
  \bibinfo{number}{4} (\bibinfo{date}{April} \bibinfo{year}{2009}),
  \bibinfo{pages}{65–76}.
\newblock
\showISSN{0001-0782}
\href{https://doi.org/10.1145/1498765.1498785}{doi:\nolinkurl{10.1145/1498765.1498785}}


\bibitem[Won et~al\mbox{.}(2023)]%
        {Won_2023}
\bibfield{author}{\bibinfo{person}{William Won}, \bibinfo{person}{Taekyung
  Heo}, \bibinfo{person}{Saeed Rashidi}, \bibinfo{person}{Srinivas Sridharan},
  \bibinfo{person}{Sudarshan Srinivasan}, {and} \bibinfo{person}{Tushar
  Krishna}.} \bibinfo{year}{2023}\natexlab{}.
\newblock \showarticletitle{ASTRA-sim2.0: Modeling Hierarchical Networks and
  Disaggregated Systems for Large-model Training at Scale}. In
  \bibinfo{booktitle}{\emph{2023 IEEE International Symposium on Performance
  Analysis of Systems and Software (ISPASS)}}. \bibinfo{publisher}{IEEE},
  \bibinfo{pages}{283–294}.
\newblock
\href{https://doi.org/10.1109/ispass57527.2023.00035}{doi:\nolinkurl{10.1109/ispass57527.2023.00035}}


\bibitem[Xiang et~al\mbox{.}(2026)]%
        {xiang2025servegen}
\bibfield{author}{\bibinfo{person}{Yuxing Xiang}, \bibinfo{person}{Xue Li},
  \bibinfo{person}{Kun Qian}, \bibinfo{person}{Yan Zhang},
  \bibinfo{person}{Wenyuan Yu}, \bibinfo{person}{Ennan Zhai},
  \bibinfo{person}{Xin Jin}, {and} \bibinfo{person}{Jingren Zhou}.}
  \bibinfo{year}{2026}\natexlab{}.
\newblock \showarticletitle{{ServeGen}: Workload Characterization and
  Generation of Large Language Model Serving in Production}. In
  \bibinfo{booktitle}{\emph{23rd USENIX Symposium on Networked Systems Design
  and Implementation (NSDI 26)}}. \bibinfo{publisher}{USENIX Association},
  \bibinfo{address}{Renton, WA}, \bibinfo{pages}{1845--1859}.
\newblock
\showISBNx{978-1-939133-54-0}
\urldef\tempurl%
\url{https://www.usenix.org/conference/nsdi26/presentation/xiang-servegen}
\showURL{%
\tempurl}


\bibitem[Xu et~al\mbox{.}(2026)]%
        {xu2026aiconfigurator}
\bibfield{author}{\bibinfo{person}{Tianhao Xu}, \bibinfo{person}{Yiming Liu},
  \bibinfo{person}{Xianglong Lu}, \bibinfo{person}{Yijia Zhao},
  \bibinfo{person}{Xuting Zhou}, \bibinfo{person}{Aichen Feng},
  \bibinfo{person}{Yiyi Chen}, \bibinfo{person}{Yi Shen}, \bibinfo{person}{Qin
  Zhou}, \bibinfo{person}{Xumeng Chen}, \bibinfo{person}{Ilya Sherstyuk},
  \bibinfo{person}{Haorui Li}, \bibinfo{person}{Rishi Thakkar},
  \bibinfo{person}{Ben Hamm}, \bibinfo{person}{Yuanzhe Li},
  \bibinfo{person}{Xue Huang}, \bibinfo{person}{Wenpeng Wu},
  \bibinfo{person}{Anish Shanbhag}, \bibinfo{person}{Harry Kim},
  \bibinfo{person}{Chuan Chen}, {and} \bibinfo{person}{Junjie Lai}.}
  \bibinfo{year}{2026}\natexlab{}.
\newblock \bibinfo{title}{AIConfigurator: Lightning-Fast Configuration
  Optimization for Multi-Framework LLM Serving}.
\newblock
\showeprint[arxiv]{2601.06288}~[cs.LG]
\urldef\tempurl%
\url{https://arxiv.org/abs/2601.06288}
\showURL{%
\tempurl}


\bibitem[Yu et~al\mbox{.}(2022)]%
        {yu2022orca}
\bibfield{author}{\bibinfo{person}{Gyeong-In Yu}, \bibinfo{person}{Joo~Seong
  Jeong}, \bibinfo{person}{Geon-Woo Kim}, \bibinfo{person}{Soojeong Kim}, {and}
  \bibinfo{person}{Byung-Gon Chun}.} \bibinfo{year}{2022}\natexlab{}.
\newblock \showarticletitle{Orca: A Distributed Serving System for
  {Transformer-Based} Generative Models}. In \bibinfo{booktitle}{\emph{16th
  USENIX Symposium on Operating Systems Design and Implementation (OSDI 22)}}.
  \bibinfo{publisher}{USENIX Association}, \bibinfo{address}{Carlsbad, CA},
  \bibinfo{pages}{521--538}.
\newblock
\showISBNx{978-1-939133-28-1}
\urldef\tempurl%
\url{https://www.usenix.org/conference/osdi22/presentation/yu}
\showURL{%
\tempurl}


\bibitem[Yu et~al\mbox{.}(2026)]%
        {yu2025voltana}
\bibfield{author}{\bibinfo{person}{Jiahuan Yu}, \bibinfo{person}{Aryan Taneja},
  \bibinfo{person}{Junfeng Lin}, {and} \bibinfo{person}{Minjia Zhang}.}
  \bibinfo{year}{2026}\natexlab{}.
\newblock \showarticletitle{VoltanaLLM: Energy-Efficient and SLO-Aware
  Disaggregated LLM Serving via Adaptive Frequency Control Íand State-Space
  Routing}. In \bibinfo{booktitle}{\emph{ISC High Performance 2026 Research
  Paper Proceedings (41st International Conference)}}. \bibinfo{pages}{1--19}.
\newblock
\href{https://doi.org/10.23919/ISC.2026.11520495}{doi:\nolinkurl{10.23919/ISC.2026.11520495}}


\bibitem[Zhang et~al\mbox{.}(2025)]%
        {zhang2024llmcompass}
\bibfield{author}{\bibinfo{person}{Hengrui Zhang}, \bibinfo{person}{August
  Ning}, \bibinfo{person}{Rohan~Baskar Prabhakar}, {and} \bibinfo{person}{David
  Wentzlaff}.} \bibinfo{year}{2025}\natexlab{}.
\newblock \showarticletitle{LLMCompass: Enabling Efficient Hardware Design for
  Large Language Model Inference}. In \bibinfo{booktitle}{\emph{Proceedings of
  the 51st Annual International Symposium on Computer Architecture}} (Buenos
  Aires, Argentina) \emph{(\bibinfo{series}{ISCA '24})}.
  \bibinfo{publisher}{IEEE Press}, \bibinfo{pages}{1080–1096}.
\newblock
\showISBNx{9798350326581}
\href{https://doi.org/10.1109/ISCA59077.2024.00082}{doi:\nolinkurl{10.1109/ISCA59077.2024.00082}}


\bibitem[Zhong et~al\mbox{.}(2024)]%
        {zhong2024distserve}
\bibfield{author}{\bibinfo{person}{Yinmin Zhong}, \bibinfo{person}{Shengyu
  Liu}, \bibinfo{person}{Junda Chen}, \bibinfo{person}{Jianbo Hu},
  \bibinfo{person}{Yibo Zhu}, \bibinfo{person}{Xuanzhe Liu},
  \bibinfo{person}{Xin Jin}, {and} \bibinfo{person}{Hao Zhang}.}
  \bibinfo{year}{2024}\natexlab{}.
\newblock \showarticletitle{DistServe: disaggregating prefill and decoding for
  goodput-optimized large language model serving}. In
  \bibinfo{booktitle}{\emph{Proceedings of the 18th USENIX Conference on
  Operating Systems Design and Implementation}} (Santa Clara, CA, USA)
  \emph{(\bibinfo{series}{OSDI'24})}. \bibinfo{publisher}{USENIX Association},
  \bibinfo{address}{USA}, Article \bibinfo{articleno}{11},
  \bibinfo{numpages}{18}~pages.
\newblock
\showISBNx{978-1-939133-40-3}


\end{thebibliography}
